\documentclass[iop]{emulateapj}
\usepackage{graphicx}
\usepackage{amssymb}
\usepackage{amsmath}
\usepackage[backref,breaklinks,colorlinks,citecolor=blue, linkcolor=blue, urlcolor=blue]{hyperref}
\usepackage[pass]{geometry}
\usepackage{enumerate}
\usepackage{enumitem}
\usepackage{color}
\usepackage{units}

\newcommand\blfootnote[1]{
\begingroup
\renewcommand\thefootnote{}\footnote{#1}
\addtocounter{footnote}{-1}
\endgroup
}

\definecolor{update}{rgb}{1,0,0}

\makeatletter
\newcommand\footnoteref[1]{\protected@xdef\@thefnmark{\ref{#1}}\@footnotemark}
\makeatother
\renewcommand{\thempfootnote}{\arabic{mpfootnote}}

\shorttitle{KCS III: Fundamental plane of JKCS 041}
\shortauthors{L. J. Prichard et al.}

\begin{document}

%-----------------
%TITLE AND AUTHORS
%-----------------
\title{The KMOS Cluster Survey (KCS) III: fundamental plane of cluster galaxies at $\MakeLowercase{z} \simeq 1.80$ in JKCS 041\altaffilmark{*}}
\altaffiltext{*}{Based on observations obtained at the Very Large Telescope (VLT) of the European Southern Observatory (ESO), Paranal, Chile (ESO program IDs: 095.A-0137(A) and 096.A-0189(A)).}

\author{Laura J. Prichard\altaffilmark{1}$^{\star}$, Roger L. Davies\altaffilmark{1}, Alessandra Beifiori\altaffilmark{2,}\altaffilmark{3}, Jeffrey C. C. Chan\altaffilmark{2,}\altaffilmark{3,}\altaffilmark{4},  Michele Cappellari\altaffilmark{1}, Ryan C. W. Houghton\altaffilmark{1},  J. Trevor Mendel\altaffilmark{3,}\altaffilmark{2},  Ralf Bender\altaffilmark{2,}\altaffilmark{3}, Audrey Galametz\altaffilmark{3,}\altaffilmark{2}, Roberto P. Saglia\altaffilmark{3,}\altaffilmark{2}, John P. Stott\altaffilmark{1,}\altaffilmark{6}, David J. Wilman\altaffilmark{2,}\altaffilmark{3}, Ian J. Lewis\altaffilmark{1}, Ray Sharples\altaffilmark{5}, and Michael Wegner\altaffilmark{2}}
\altaffiltext{1}{Sub-Department of Astrophysics, Department of Physics, University of Oxford, Denys Wilkinson Building, Keble Road, Oxford OX1 3RH, UK}
\altaffiltext{2}{Universit{\"a}ts-Sternwarte M{\"u}nchen, Scheinerstra{\ss}e 1, D-81679 M{\"u}nchen, Germany}
\altaffiltext{3}{Max-Planck-Institut f{\"u}r Extraterrestrische Physik, Giessenbachstra{\ss}e 1, D-85748 Garching, Germany}
\altaffiltext{4}{Department of Physics and Astronomy, University of California, Riverside, CA 92521, USA}
\altaffiltext{5}{Centre for Advanced Instrumentation, Department of Physics, Durham University, South Road, Durham DH1 3LE, UK}
\altaffiltext{6}{Department of Physics, Lancaster University, Lancaster LA1 4YB, UK}

%\date*{Accepted October 26, 2017.}
%\pubyear{2017}

%----------------
%BEGIN DOC
%----------------
%----------------
%ABSTRACT
%----------------
\begin{abstract}
We present data for 16 galaxies in the overdensity JKCS 041 at $z \simeq 1.80$ as part of the K-band Multi-Object Spectrograph (KMOS) Cluster Survey (KCS). With 20-hour integrations, we have obtained deep absorption-line spectra from which we derived velocity dispersions for seven quiescent galaxies. We combined photometric parameters derived from \textit{Hubble Space Telescope} images with the dispersions to construct a fundamental plane (FP) for quiescent galaxies in JKCS 041. From the zero-point evolution of the FP, we derived a formation redshift for the galaxies of $z_{form} = 3.0\pm0.3$, corresponding to a mean age of $1.4\pm0.2$ Gyrs. We tested the effect of structural and velocity dispersion evolution on our FP zero point and found a negligible contribution when using dynamical mass-normalized parameters ($\sim 3\%$), but a significant contribution from stellar-mass-normalized parameters ($\sim 42 \%$). From the relative velocities of the galaxies, we probed the three-dimensional structure of these 16 confirmed members of JKCS 041, and found that a group of galaxies in the south west of the overdensity had systematically higher velocities. We derived ages for the galaxies in the different groups from the FP. We found the east-extending group had typically older galaxies ($2.1\substack{+0.3\\-0.2}$ Gyrs), than those in the south-west group ($0.3\pm0.2$ Gyrs). Although based on small numbers, the overdensity dynamics, morphology, and age results could indicate that JKCS 041 is in formation and may comprise two merging groups of galaxies. The result could link large-scale structure to ages of galaxies for the first time at this redshift.
\end{abstract}


%-----------------------
%INTRODUCTION
%-----------------------
\section{Introduction}
\label{sec:intro}

\blfootnote{\vspace{-4 mm}$^{\star}$Email: \href{mailto:Laura.Prichard@physics.ox.ac.uk}{Laura.Prichard$@$physics.ox.ac.uk}}

In the hierarchical formation model of $\Lambda$ cold dark matter ($\Lambda$CDM) cosmology, the largest scale structures, such as clusters of galaxies, are the last to be formed \citep[e.g.,][]{Press1974, Toomre1977, White1978}. Therefore, as we go to higher redshifts, galaxy clusters are increasingly rare \citep[e.g.,][]{Haiman2001}. Currently, only a handful of rich overdensities, with a well-defined `red sequence' of galaxies -- a narrow feature on a galaxy color-magnitude diagram (CMD) -- exist at $1.5 \lesssim z \lesssim 2.5$ \citep[][see a recent review by \citealt{Overzier2016}]{Kurk2009, Papovich2010, Tanaka2010, Gobat2011, Gobat2013, Santos2011, Stanford2012, Zeimann2012, Muzzin2013a, Andreon2014, Newman2014, Wang2016}.

The evolution of galaxies is connected to their environment; the morphology-density relation \citep{Dressler1980} describes the phenomenon whereby `red and dead' elliptical galaxies are more common in denser environments, i.e., towards the center of galaxy clusters. Therefore, processes that occur within these dense environments alter the evolutionary paths of their member galaxies. Probing galaxy evolution in dense environments out to increasingly higher redshift has long been a key focus of extragalactic astronomy \citep[e.g.,][]{Dressler1997, Stanford1998, Poggianti1999, Postman2005}. However, capturing the evolutionary processes within distant quiescent galaxies out to $z \sim 2$ has so far proved challenging for all but the most massive galaxies. Rare, massive, high-redshift overdensities provide a valuable opportunity to observe the largest samples of early-type galaxies (ETGs) out to $z \sim 2$ in order to constrain the evolution of galaxies at these redshifts by studying their stellar light.

Scaling relations of ETGs provide powerful diagnostic tools for probing the evolution of galaxies. One such scaling relation is the fundamental plane \citep[FP;][]{Djorgovski1987, Dressler1987, Jorgensen1996}. The FP is the relationship between size, surface brightness, and velocity dispersion. In the nearby Universe, the existence of the FP was shown to be due almost entirely to virial equilibrium combined with a systematic variation of the mass-to-light ratio \citep[$M/L$;][]{Cappellari2006, Cappellari2013a, Bolton2007, Auger2010b}. See \cite{Cappellari2016} for a recent review. 

The FP zero-point is known to evolve strongly with redshift and can be used to determine the stellar ages of galaxies \citep[e.g.][]{Beifiori2017}. Assuming ETGs are a homologous population, this zero-point evolution can be explained by the $M/L$ evolution of the aging stellar population \citep[e.g.,][]{vanDokkum1996, vanDokkum2007, Holden2010}, or alternatively by size evolution \citep{Saglia2010, Saglia2016}. To construct an FP, deep continuum spectroscopy is required to obtain accurate absorption-line stellar velocity dispersions. Given the difficulties in obtaining deep enough spectra to measure reliable velocity dispersions for individual galaxies, FPs have only been constructed out to $z \sim 2$ comprising the brightest (typically the central) galaxies. These studies have shown that the FP holds to $z \sim$ 2 \citep{Toft2012, Bezanson2013b, vdSande2014}.

Performing these types of studies up to $z \sim 2$ has proved costly in telescope time. However, absorption line-studies of galaxies are essential for constraining the kinematics and properties of the stellar content in galaxies. The first detection of absorption lines at $z > 2$ came from a 29-hour integration of one galaxy at $z = 2.2$ \citep{Kriek2009}; a high velocity dispersion ($\sigma = 510\substack{+165 \\ -95}$ km s$^{-1}$) was also determined for this galaxy \citep{vanDokkum2009}. Even with extensive efforts going to expand this sample of dispersion values at $z \sim 2$, only around a dozen measurements have been obtained directly for individual galaxies at these redshifts \citep{vanDokkum2009, Onodera2010, Toft2012, vdSande2013, Belli2014b, Belli2017}. A few more have been determined for lensed ETGs at $z > 2$, which negates the need for such long integrations, but these cases are rare \citep[e.g.,][]{Newman2015, Hill2016, Toft2017}. 

Multi-object spectrographs can improve observing efficiency of high-$z$ galaxies, provided target quiescent galaxies have high spatial density. High redshift galaxy overdensities with a well-populated red sequence satisfy these constraints. The K-band Multi-Object Spectrograph \citep[KMOS;][]{Sharples2013} on the Very Large Telescope (VLT) in Chile is a near-infrared (NIR) multi-integral field unit (IFU) instrument. The IFUs of KMOS were specifically designed to match galaxy sizes at high redshift, and 24 separate targets can be observed simultaneously, vastly improving efficiency of IFU observations. At $1 < z < 2$, the rest-frame optical region lies in the NIR range of KMOS. Emission and absorption lines in the rest-frame optical are valuable tools for probing the stellar populations and kinematics within galaxies. However, absorption-line studies at this distance are technically challenging, requiring high signal-to-noise (S/N) that can only be achieved from long integration times. 

As part of the KMOS Guaranteed Time Observations (GTO) project, the KMOS Cluster Survey \citep[KCS;][Davies, Bender et al., in prep.]{Davies2015}, ETGs in overdensities between $1 < z < 2$ were observed for 20 hours on source, to constrain galaxy evolutionary processes in dense environments at higher redshifts than previously possible \citep[see][for a summary]{Beifiori2017}. The redshift range covers the peak epoch of star formation \citep{Madau1996,Madau2014}, around 10 Gyrs ago, an important phase in galaxy evolution when their stellar populations were being established. Coupling KMOS data with \textit{Hubble Space Telescope} (\textit{HST}) imaging, the size, age, morphology, and star formation histories (SFHs) of the galaxies can be investigated \citep{Chan2016, Beifiori2017, Chan2017}.

The highest redshift KCS target, JKCS 041, is an ETG-rich, $z \simeq 1.80$ overdensity. JKCS 041 was first identified by \cite{Andreon2009}, when detected with diffuse X-ray \textit{Chandra} observations. The overdensity was spectroscopically confirmed with 19 members, and three candidate members, at $z = 1.803$ with \textit{HST} grism spectroscopy \citep{Newman2014}. The total overdensity mass was determined to be in the range $\log(M/M_{\odot}) = 14.2 - 14.5$ \citep{Andreon2014}. Further investigation of the overdensity showed that the mass-matched field sample at the same redshift were not as quiescent \citep{Newman2014}, providing compelling evidence for environmental quenching. JKCS 041 was therefore an ideal system to target in order to improve our understanding of galaxy evolution out to $z \sim 2$.
	
In this paper we investigated the properties of the highest redshift overdensity in the KCS sample, JKCS 041 at $z \simeq 1.80$. We will present the KMOS spectroscopic data of the galaxies, which when combined with \textit{HST} imaging \citep[presented in][]{Newman2014}, enabled us to construct an FP of galaxies in JKCS 041 in order to determine their mean stellar age. We used dynamical information to construct a three-dimensional view of the observed overdensity members. This paper is organized as follows, an overview of KCS is in Sec. \ref{sec:kcs}. The sample selection is covered in Sec. \ref{sec:sampsel}. The reduction and analysis of \textit{HST} images are described in Sec. \ref{sec:phot}. The KMOS observations, data reduction, and analysis of the spectroscopic sample are described in Sec. \ref{sec:spec}. The FP and derivation of mean galaxy ages are presented in Sec. \ref{sec:fp}. A discussion of the 3D structure of the overdensity is in Sec. \ref{sec:struct}. We discuss our results from JKCS 041 in the context of the current literature in Sec. \ref{sec:disc}. Finally, we present a summary of our findings in Sec. \ref{sec:conc}. Throughout the paper, we assume $\Lambda$CDM cosmology with $\Omega_m = 0.3$, $\Omega_{\Lambda} = 0.7$, $H_0 = 70$ km s$^{-1}$ Mpc$^{-1}$ (these agree well with the latest results from \citealt{Planck2016}), and use the AB magnitude system \citep{Oke1983}.

\section{The KMOS Cluster Survey}
\label{sec:kcs}

The aim of KCS was to constrain galaxy properties in dense environments out to $z \sim 2$ by studying stellar kinematics and the evolution of stellar populations. See \cite{Beifiori2017} for a detailed description of the selection function for galaxies in the KCS sample, and for additional details of KCS that we will summarize here. 

The target overdensities for KCS were required to be between $1 \lesssim z \lesssim 2$, have many bright red-sequence galaxies, and multi-band \textit{HST} imaging. Ideally each overdensity contained $> 20$ sufficiently bright quiescent galaxies on the red sequence to make best use of the 24 IFUs of KMOS. We required that each target had multi-band \textit{HST} imaging as it was used in both the reduction of the KMOS spectra and to investigate the photometric and structural properties of the galaxies within the overdensities. Of the four main KCS overdensities (three in \citealt{Beifiori2017}, and JKCS 041 in this paper), $\sim$ 100 galaxies were observed and $\sim 70$ of these were quiescent galaxies. For the three KCS overdensities in \cite{Beifiori2017}, quiescent galaxies were selected to lie within 2$\sigma$ of the fitted red sequence. The brightest red-sequence galaxies with spectroscopic redshift measurements in the literature were prioritized, followed by red-sequence galaxies with no redshift values available, then fainter red-sequence or emission-line galaxies were observed to fill remaining IFUs in the field of view. These quiescent galaxies were observed for $\sim$ 20 hours on source in order to get down to the magnitudes and sensitivity needed to measure absorption lines. 

In \cite{Beifiori2017}, we measured stellar velocity dispersions and combined them with the photometric properties derived from the multi-band \textit{HST} imaging presented in \cite{Chan2016, Chan2017}. These values were used to construct FPs and to derive population ages for the galaxies from the zero-point evolution for three KCS overdensities: XMMU J2235.3-2557 at $z = 1.39$ \citep[XMM2235;][]{Mullis2005, Rosati2009}, XMMXCS J2215.9-1738 at $z = 1.46$ \citep[XMM2215;][]{Stanford2006, Hilton2007, Hilton2009, Hilton2010}, and Cl 0332-2742 at $z = 1.61$ \citep[Cl0332;][]{Kurk2009, Castellano2007}. Ages derived from the full-spectral fitting of the galaxy spectra in these overdensities will be covered in an upcoming paper \citep{Houghton2017}.

The analysis presented in this paper builds upon the work described in \cite{Beifiori2017}. Using FP analysis, \cite{Beifiori2017} found ages of the high mass ($\log (M_*/M_{\odot}) > 11$) galaxies in Cl0332, XMM2215, and XMM2235, of $1.20\substack{+1.03\\-0.47}$ Gyrs, $1.59\substack{+1.40\\-0.62}$ Gyrs, and $2.33\substack{+0.86\\-0.51}$ Gyrs respectively. The results showed that the galaxies in the three overdensities were consistent with passive evolution and had formation epochs consistent within errors. Although interestingly, for XMM2235, the more relaxed and massive cluster in the sample, \cite{Beifiori2017} found a hint of an older relative formation age for the most massive galaxies. This could imply that galaxies in a more virialized, relaxed environment (i.e., XMM2235) undergo accelerated evolution, as found in previous studies \citep[e.g.,][]{Gebhardt2003, Saglia2010}. We extend the work of \cite{Beifiori2017} in this paper to investigate the ages of the galaxies in JKCS 041 at $z \simeq 1.80$ through analysis of the FP.

\begin{figure} 
\includegraphics[width=0.5\textwidth , trim=28 18 25 60,clip]{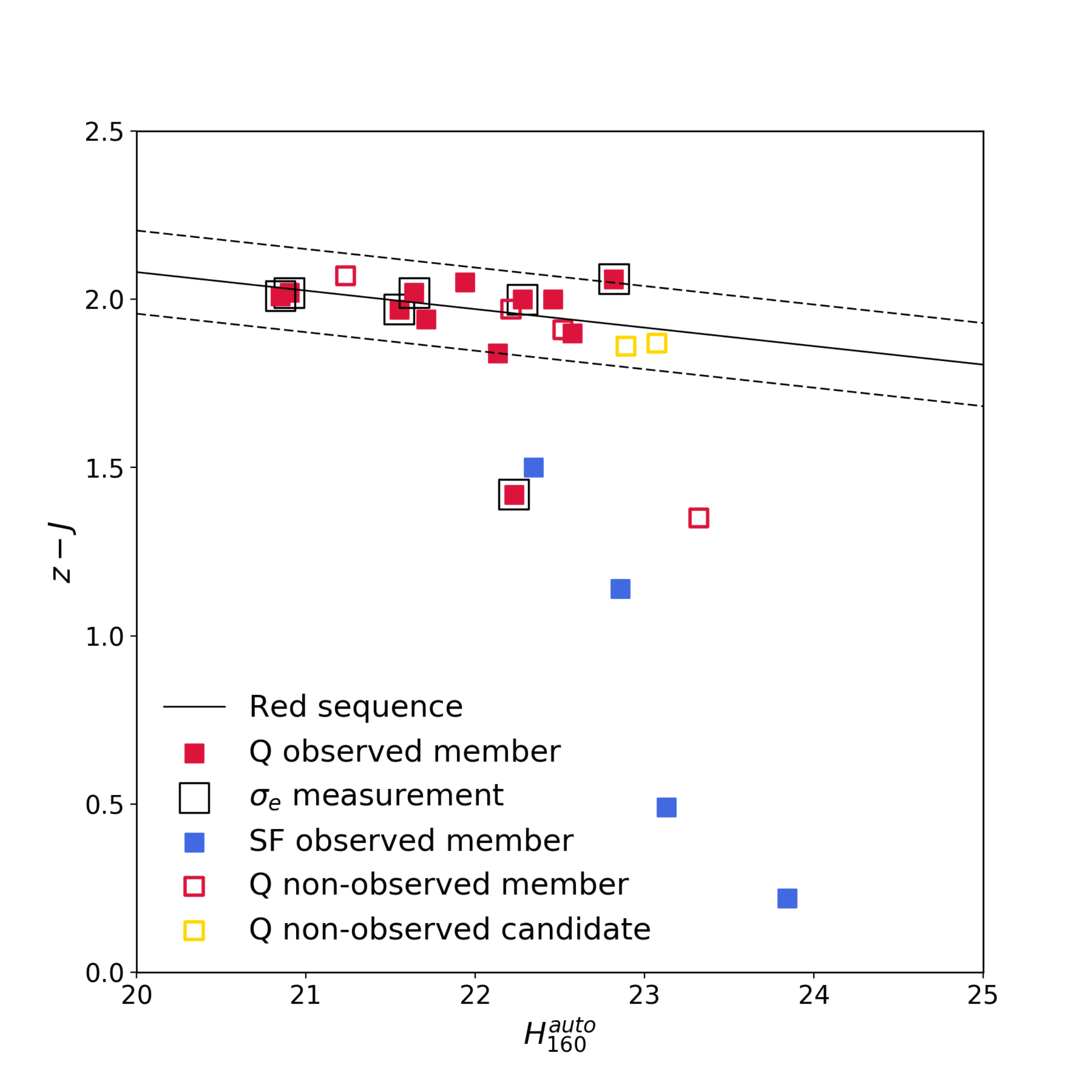}\caption[]{Galaxy CMD for confirmed and candidate members of JKCS 041. The quiescent (Q; red) and star forming (SF; blue) confirmed member galaxies that have been observed with KMOS (filled squares), and confirmed but not observed (unfilled), are shown along with the unobserved candidate members (unfilled yellow squares; see Sec. \ref{sec:targ}). Those objects for which we derived velocity dispersion measurements are shown by unfilled black squares (see Sec. \ref{sec:kinfit}). The best fit to the confirmed red-sequence galaxies is shown by the solid line, while the dashed lines shows the 2$\sigma$ scatter. The $z-J$ color \citep{Newman2014} is plotted against our $H$-band \textsc{SExtractor} total magnitude estimate \texttt{MAG\_AUTO} ($H^{auto}_{160}$; see Sec. \ref{sec:photana}). See Sec. \ref{sec:sampsel}.\small}
\label{fig:cmd}
\end{figure}

\section{Sample Selection}
\label{sec:sampsel}

JKCS 041 is the highest redshift overdensity in the KCS sample at $z \simeq 1.80$. When spectroscopically confirmed by \cite{Newman2014}, JKCS 041 had 19 confirmed members, 15 of which were quiescent. \cite{Newman2014} determined photometric redshift (photo-$z$) measurements from multi-band photometry, and derived grism redshifts for all objects where possible. \cite{Newman2014} then constructed a CMD for all objects in the field at limiting magnitudes in the F160W band of \textit{HST} ($H_{160}$) $< 25.5$ for emission line objects, and $H_{160} <$ 23.3 for non-emission line `continuum' objects. Three galaxies were on the red sequence that were not confirmed from grism redshifts (due to contamination) to be either members of JKCS 041 (at $z \sim 1.80$) or not in the overdensity; these were flagged as candidate overdensity members. We observed one of these candidate members as part of our KMOS spectroscopic sample (ID 772) and confirmed its membership, bringing the total number of confirmed galaxies in JKCS 041 to 20.  

The galaxy CMD shown in Fig. \ref{fig:cmd} is composed of $z-J$ values from ground-based photometry from \cite{Newman2014} and our $H_{160}$ total magnitudes derived using the  \textsc{SExtractor} \texttt{MAG\_AUTO} estimate ($H^{\rm auto}_{160}$; see Sec. \ref{sec:photana}). In Fig. \ref{fig:cmd}, we show the members we observed with KMOS (filled squares), those we did not (unfilled), and indicate whether the galaxies are star forming (SF; blue) or quiescent (Q; red). We also include the two red-sequence selected candidate members of JKCS 041 that we did not observe (unfilled yellow squares). We show the best fit from a least-squares method to the confirmed red-sequence galaxies (solid line), and indicate the 2$\sigma$ scatter in the red sequence (dashed lines). Also highlighted are the seven galaxies for which we derive velocity dispersion measurements (see Sec. \ref{sec:kinfit}). JKCS 041 has a well-defined red sequence with little scatter, setting it apart in maturity from most overdensities at comparable redshifts.

\begin{figure*} 
\centering
\includegraphics[width=0.7\textwidth , trim=30 10 0 50,clip]{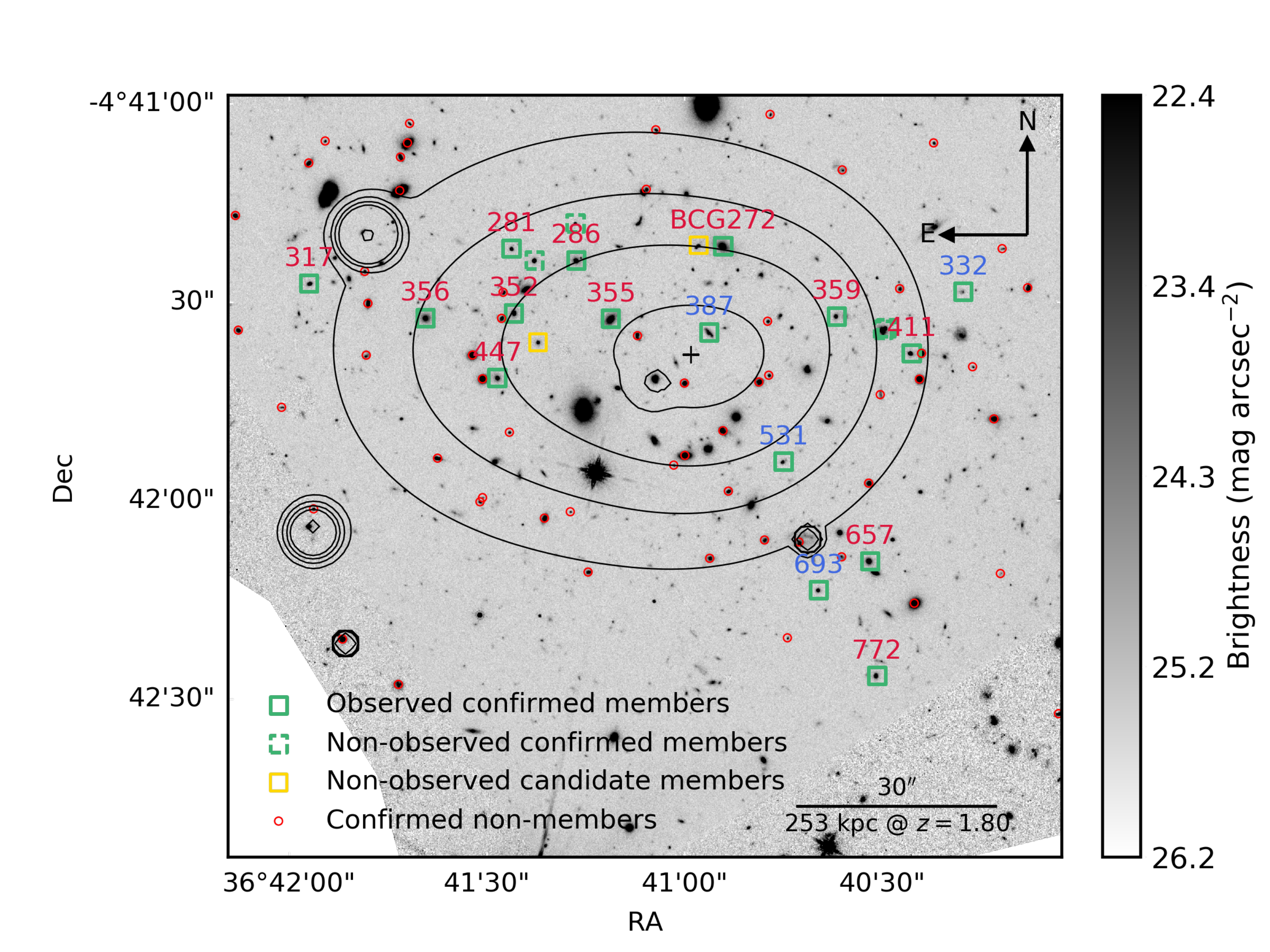}
\caption[]{\textit{HST} $H_{160}$ band image of JKCS 041. The observed overdensity members (solid green squares) are shown with their IDs color coded for whether they are SF (blue) or quiescent (red). The other spectroscopically confirmed members that were not observed with KMOS (dashed green squares), and the unconfirmed candidate members on the red sequence that we did not observe (yellow squares) are all quiescent \citep{Newman2014}. The X-ray contours are from \cite{Andreon2009}, and the geometric center of the galaxies is shown with a `+'. All objects confirmed as non-overdensity members with grism redshifts \citep{Newman2014} are shown (small red circles; see Sec. \ref{sec:struct} for more details). See Sec. \ref{sec:targ}.\small}
\label{fig:hst}
\end{figure*}

Selection effects for JKCS 041 as compared to the other KCS overdensities, are difficult to quantify. We were limited in our selection of targets by the density of the confirmed members of JKCS 041. The sample could also be limited by the size of the \textit{HST} Wide Field Camera 3 (WFC3) field of view, these effects would then also be present in the sample selection from \cite{Newman2014}. As discussed in \cite{Newman2014}, the `continuum' sample of mostly quiescent galaxies in the WFC3 image for which they derived spectrophotometric redshifts was strictly flux limited ($H_{160} <$ 23.3 mag). \cite{Newman2014} found that this was an almost mass-complete sample, with 88$\%$ completeness at $\log(M_{*}/M_{\odot}) >$ 10.6 for this magnitude limit at $z = 1.80$. This mass completeness estimate was based on a larger area sample from \cite{Newman2012} for this redshift. The two remaining candidate members have $\log(M_{*}/M_{\odot}) <$ 11, so it is likely that the overdensity members are complete above this mass limit within $R_{500}$ (which is within the WFC3 footprint, see Fig. 3 of \citealt{Newman2014}) of the overdensity center. For the remaining 10.6 $< \log(M_{*}/M_{\odot}) <$ 11 galaxies, accounting for the two remaining candidate cluster members, the sample is $\sim 78\%$ complete. \cite{Newman2014} explained that the completeness for the emission-line sample was harder to quantify. 

With so few confirmed member galaxies not observed with KMOS (four) from this parent sample of 20 members, it is difficult to robustly quantify the selection effects of the observed sample beyond the completeness already discussed by \cite{Newman2014}. As seen on the CMD in Fig. \ref{fig:cmd}, the galaxies for which we derive velocity dispersions (unfilled black squares), fairly evenly span the range of magnitudes we sample, helping to reduce any bias (beyond these completeness limits) in the ages we determined from them. 

\subsection{KMOS target selection}
\label{sec:targ}

As shown in Fig. \ref{fig:hst}, the overdensity spans $\sim 1.9\arcmin \times 1.3\arcmin$ (as determined from the extent of the galaxies), which at $z = 1.80$ is $\sim 1.0 \times 0.6$ Mpc in size. Given the relatively tight morphology of the 20 confirmed members compared to the patrol region of KMOS ($7\arcmin$ diameter), the number of galaxies that could be observed with KMOS was limited by the proximity constraint of the IFUs. Each IFU is $2.8\arcsec \times 2.8\arcsec$, 14 $\times$ 14 pixels, which at $z = 1.80$ is $\sim$ 24 kpc, and is sufficiently large to encompass $>$ the effective radius ($R_e$) of the galaxies. To determine the optimal configuration for the 24 robotic arms of KMOS, we used the KMOS ARM Allocator \citep[\textsc{karma};][]{Wegner2008} software. There are two layers of arms to prevent collisions, however, two arms within the same layer cannot come within $\sim 6 \arcsec$ (between IFU centers). 

In a dense region of four galaxies, where only two could be observed, we prioritized the brightest quiescent members (e.g. IDs 286 $\&$ 281 over IDs 289 $\&$ 255). We aimed to get the isolated (separated with respect to other members of JKCS 041 $> 6\arcsec$) sources, as unresolved pairs were harder to extract reliable 1D spectra from (e.g. IDs 375 $\&$ 376). However, we were able to extract spectra for confirmed members with close resolved neighbors (e.g. IDs 359 $\&$ 411) as their separation was larger than that of the full-width-half-maximum (FWHM) of the KMOS point spread function (PSF). As we had spare IFUs after targeting all possible quiescent members, we also observed member SF galaxies and quiescent candidate members. We selected targets in order of preference, with the final configuration made to reflect these priorities in each area of the sky:

\begin{enumerate}[leftmargin=0.3 in]
\item Quiescent, bright, isolated, confirmed overdensity members
\item SF, isolated, confirmed overdensity members
\item Paired confirmed members
\item Quiescent, candidate members
\end{enumerate}

In summary, we targeted 16 galaxies, 15 of these were spectroscopically confirmed, and one was a candidate member, however with KMOS observations we confirmed it as a member. Of the 16 galaxies observed with KMOS, 12 of these were quiescent (including ID 772, the candidate member that we spectroscopically confirmed), and four were SF \citep[as classified on a $UVJ$ diagram by][]{Newman2014}. The \textit{HST} image in Fig. \ref{fig:hst} indicates those confirmed overdensity members observed with KMOS (solid green squares), showing their IDs color coded for whether they are quiescent (red) or SF (blue). The green dashed squares indicate the remaining confirmed members not observed with KMOS (four galaxies, all quiescent), due to their close proximity to other targets. We then also show the remaining two candidate cluster members (yellow squares), as those not spectroscopically confirmed by \cite{Newman2014} or observed with KMOS, but that lie on the red sequence.

%---------------------
%PHOT DATA
%---------------------
\section{Photometry}
\label{sec:phot}

\subsection{Photometric data}
\label{sec:photdata}

The multi-band \textit{HST} images used in this paper were first presented in \cite{Newman2014}. Images for JKCS 041 were taken using the infrared filters F160W ($H_{160}$ band) and F105W ($Y_{105}$ band) on WFC3 (GO 12927, Cycle 20, P.I. Newman). A four-point dither pattern was used, and the images were combined with grism pre-images, giving total exposure times in each band of $\sim$ 4.5~ks in $H_{160}$ and $\sim$ 2.7~ks in $Y_{105}$. For a self-consistent KCS catalogue, we re-reduced and analyzed the images presented in \cite{Newman2014}. 

\subsection{Reduction of photometry}
\label{sec:photred}

To reduce the \textit{HST} images of JKCS 041 for this paper, we used \textsc{astrodrizzle} from \textsc{DrizzlePac} \citep[version 2.0;][]{Gonzaga2012}, which is an updated version of the \textsc{multidrizzle} software \citep{Koekemoer2002}. Following a similar technique to \cite{Chan2016, Chan2017}, for each band we used the routine \textsc{tweakreg} to align the images on to a common reference frame. The aligned frames were then drizzled together onto a pixel scale of $0\farcs06$. We then matched the World Coordinate System of the combined images in each band to that used in \cite{Newman2014}. For continuity between studies, we also adopted the galaxy IDs and $UVJ$ classifications from \cite{Newman2014} to distinguish between SF and quiescent galaxies. 

\begin{figure*}
\centering 
\includegraphics[width=0.85\textwidth , trim=50 100 50 110,clip]{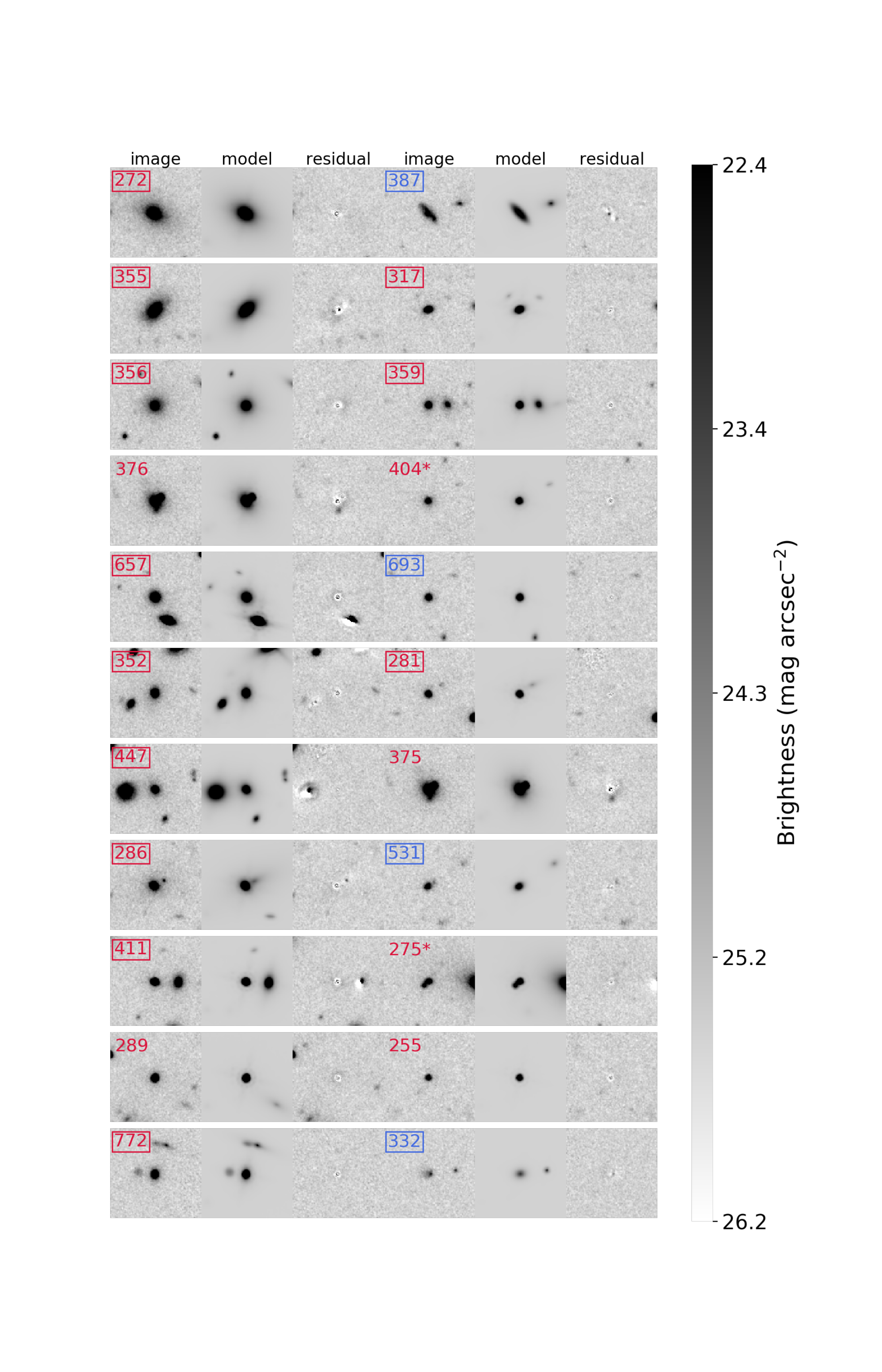}
\caption[]{$H_{160}$ postage stamps of all the confirmed and candidate members of JKCS 041. We indicate whether they are quiescent (red ID), SF (blue ID), observed with KMOS (boxed IDs), not observed members, or not observed candidate members (*). For each galaxy (from left to right) we plot the $H_{160}$-band \textit{HST} image, the \textsc{galapagos-2.2.5b} fit to the central target galaxy and any other simultaneously fitted sources that may affect the fit, and the residual. The galaxies are ordered in columns by total integrated magnitudes as derived from S{\'e}rsic fits ($H^{tot}_{160}$); brightest (BCG, ID 272) to faintest. Each postage stamp is $6.78\arcsec \times 6.54\arcsec$. See Sec. \ref{sec:photana}.\small}
\label{fig:photom}
\end{figure*}

\subsection{Photometric analysis}
\label{sec:photana}

To analyze the photometry, we fitted 2D S{\'e}rsic profiles to the galaxies in the images to extract their light-weighted properties. \cite{Chan2016, Chan2017} performed extensive photometric analysis on the other overdensities in the KCS sample: we adopted the same methods for JKCS 041. In \cite{Chan2016, Chan2017}, light-weighted structural parameters were obtained for the galaxies in the \textit{HST} images using a modified version of the Galaxy Analysis over Large Areas: Parameter Assessment by GALFITting Objects from SExtractor \citep[\textsc{galapagos};][]{Barden2012} software (version 1.0), that utilizes \textsc{galfit} \citep[version 3.0.5;][]{Peng2002, Peng2010} and \textsc{SExtractor} \citep{Bertin1996}. 

For the photometric analysis of the galaxies in JKCS 041, we used an updated and currently maintained version of \textsc{galapagos} \citep[version 2.2.5b\footnote{Available from Boris H{\"a}u{\ss}ler's GitHub page: \href{https://github.com/MegaMorph/galapagos}{https://github.com/MegaMorph/galapagos}.};][]{Haussler2013}. \textsc{galapagos-2} and above uses an adapted version of \textsc{galfit3} \citep{Peng2010}, designed to fit multiple bands simultaneously, called \textsc{galfitm} \citep{Haussler2013}. Although running the bands separately in each filter to maintain as much consistency as possible between this work and that of \cite{Chan2016, Chan2017}, we used \textsc{galfitm} (version 1.2.1) for compatibility with \textsc{galapagos-2.2.5b}. However, \textsc{galfitm} handles single bands similarly to \textsc{galfit}. Most adaptations made to the original \textsc{galapagos} used in \cite{Chan2016, Chan2017} have now been implemented in the currently maintained version (\textsc{galapagos-2.2.5b}) used for this work, with the exception of running \textsc{SExtractor} in dual image mode. We therefore adapted \textsc{galapagos-2.2.5b} to run \textsc{SExtractor} in dual image mode with the deeper $H_{160}$ image as the detection band for the fainter $Y_{105}$ image. 

\textsc{galapagos} works by first running \textsc{SExtractor} to detect sources, and derives basic photometric properties including a total integrated magnitude estimate using its \texttt{MAG\_AUTO} parameter (shown in Fig. \ref{fig:cmd}). It then cuts out postage stamps for each of these sources using information derived from \textsc{SExtractor}. \textsc{galapagos} then estimates the sky background, using rigorous masking of sources, and a flux-growth curve method. The sky is estimated in a series of elliptical annuli, the expanse of which is not limited by the size of the postage stamp. This value of the sky is then fixed for the \textsc{galfit} fit; avoiding contamination effects when \textsc{galfit} is fitting multiple neighboring sources. We found that because the sky was fixed prior to fitting the galaxies, \textsc{galapagos} could more reliably extract sources, especially those with close neighbors, as compared to using \textsc{galfit} alone. Finally, it sets up and runs \textsc{galfit} using information obtained from the previous steps. It compiles a final output catalogue of values for each stage and for each source found by \textsc{SExtractor}. 

The PSF was derived using a similar method to \cite{Newman2012} and \cite{Chan2016, Chan2017}; we median stacked four stars from the deepest part of the image (these were the same point sources \citealt{Newman2014} stacked to make their PSF). The FWHM for the PSF was 0.15$\arcsec$ for the $H_{160}$ band. For deriving accurate aperture photometry to calculate galaxy colors ($Y_{105}-H_{160}$), we PSF matched the images by convolving the $H_{160}$-band image with the $Y_{105}$ PSF and $Y_{105}$-band image with the $H_{160}$ PSF. We then extracted colors within fixed 1$\arcsec$ diameter apertures using \textsc{SExtractor} in dual image mode with the deeper $H_{160}$ band as the detection image.

When imposing an upper limit of $n = 8$ for our analysis \citep[as in][]{Newman2014}, we found that three confirmed member galaxies reached this limit with our fitting method. We therefore chose to extend our range of S{\'e}rsic indices to $0.2 < n < 10$, as it has been reported that S{\'e}rsic indices can exist up to 10 for ETGs \citep[e.g.,][]{Caon1993, Graham1996, Kormendy2009}. The total integrated magnitudes were corrected for Galactic reddening using the \textsc{ebvpy} package\footnote{Developed for \textsc{python} by R. J. Smethurst, \href{https://github.com/rjsmethurst/ebvpy}{https://github.com/rjsmethurst/ebvpy}.} which utilizes the dust maps from \cite{Schlegel1998}, and adopting the E($B-V$) recalibration from \cite{Schlafly2011}. A full comparison of our derived values and those of \cite{Newman2014} is in Appendix \ref{sec:compphot}. In general, we found our derived parameters to be consistent to those of \cite{Newman2014} within 1$\sigma$ errors.

Fig. \ref{fig:photom} shows postage stamps of all 20 confirmed and the two candidate members of JKCS 041 (see Sec. \ref{sec:sampsel}). We show the quiescent (red IDs) and SF members (blue IDs) that we observed with KMOS (boxed IDs), confirmed members we did not observe, and unobserved candidate members (*). In two columns, we show all members in order of total integrated magnitude from S{\'e}rsic fits ($H^{tot}_{160}$); from the brightest cluster galaxy (BCG, ID 272) to the faintest. For each galaxy, we show (left to right) the input $H_{160}$-band \textit{HST} image, the \textsc{galapagos2.2.5b} model fit to the galaxy (centered) and any neighboring sources that would affect the fit, and the residual. 

\begin{deluxetable*}{ccccccccccccccc}
\tablewidth{\textwidth}    %INCLUDED
\tablecaption{Photometric properties of confirmed and candidate members of JKCS 041. \label{tab:photprops}}			%INCLUDED
\tablehead{\colhead{ID\footnotemark{\label{n14}}} & \colhead{RA} & \colhead{Dec} & \colhead{Obs.} & \colhead{Conf.} & \colhead{$UVJ$\footnoteref{n14}} & \colhead{$\log(M_*/M_{\odot})$} & \colhead{$Y_{105}-H_{160}$} & \colhead{$H^{tot}_{160}$} & \colhead{$H^{auto}_{160}$} & \colhead{$R^{maj}_{e,H_{160}}$ ($\arcsec$)} & \colhead{$R_{e,B}$ (kpc)} & \colhead{$\log \langle I_e \rangle$} & \colhead{$q$} & \colhead{$n$}}
\startdata
272 & 36.681717 & -4.689343 & 1 & 1 & Q & 11.98 & 1.52 & 20.01 & 20.90 & 1.61 & 14.38 & 2.60 & 0.69 & 8.1 \\
355 & 36.686442 & -4.692394 & 1 & 1 & Q & 11.57 & 1.29 & 20.64 & 20.85 & 0.41 & 3.70 & 3.59 & 0.52 & 3.3 \\
356 & 36.694234 & -4.692352 & 1 & 1 & Q & 11.69 & 1.51 & 20.72 & 21.55 & 1.30 & 11.63 & 2.50 & 0.97 & 9.6 \\
657 & 36.675567 & -4.702566 & 1 & 1 & Q & 11.36 & 1.47 & 21.47 & 21.64 & 0.15 & 1.31 & 4.10 & 0.88 & 3.9 \\
352 & 36.690508 & -4.692149 & 1 & 1 & Q & 11.33 & 1.58 & 21.73 & 21.94 & 0.16 & 1.38 & 3.93 & 0.69 & 4.3 \\
447 & 36.691213 & -4.694866 & 1 & 1 & Q & 10.88 & 1.11 & 21.90 & 22.23 & 0.34 & 3.06 & 3.29 & 0.78 & 4.3 \\
286 & 36.687885 & -4.689932 & 1 & 1 & Q & 11.29 & 1.66 & 21.97 & 21.71 & 0.10 & 0.91 & 4.18 & 0.62 & 3.3 \\
411 & 36.673817 & -4.693840 & 1 & 1 & Q & 11.19 & 1.57 & 22.06 & 22.13 & 0.06 & 0.51 & 4.66 & 0.52 & 1.2 \\
772 & 36.675274 & -4.707378 & 1 & 1 & Q & 11.06 & 1.42 & 22.14 & 22.28 & 0.10 & 0.90 & 4.18 & 0.58 & 3.6 \\
387 & 36.682298 & -4.692970 & 1 & 1 & SF & 10.96 & 1.36 & 22.29 & 22.34 & 0.35 & 3.12 & 3.06 & 0.32 & 0.3 \\
317 & 36.699108 & -4.690911 & 1 & 1 & Q & 10.90 & 1.30 & 22.33 & 22.46 & 0.09 & 0.79 & 4.25 & 0.39 & 1.7 \\
359 & 36.676955 & -4.692279 & 1 & 1 & Q & 10.78 & 1.21 & 22.44 & 22.57 & 0.09 & 0.79 & 4.23 & 0.84 & 6.1 \\
693 & 36.677709 & -4.703786 & 1 & 1 & SF & 10.27 & 0.92 & 22.76 & 22.86 & 0.09 & 0.79 & 4.17 & 0.90 & 1.7 \\
281 & 36.690607 & -4.689443 & 1 & 1 & Q & 10.83 & 1.45 & 22.76 & 22.82 & 0.07 & 0.61 & 4.26 & 0.65 & 0.6 \\
531 & 36.679183 & -4.698392 & 1 & 1 & SF & 9.45 & 0.40 & 22.99 & 23.13 & 0.14 & 1.28 & 3.79 & 0.58 & 1.6 \\
332 & 36.671648 & -4.691250 & 1 & 1 & SF & 8.80 & 0.14 & 23.71 & 23.84 & 0.24 & 2.18 & 3.11 & 0.70 & 1.0 \\\hline\hline\vspace{-2mm}\\
376 & 36.675004 & -4.692865 & 0 & 1 & Q & 11.72 & 1.56 & 20.74 & 21.23 & 0.58 & 5.18 & 3.18 & 0.66 & 9.1 \\
289 & 36.689651 & -4.689939 & 0 & 1 & Q & 11.06 & 1.42 & 22.11 & 22.21 & 0.06 & 0.52 & 4.66 & 0.54 & 0.7 \\
375 & 36.674881 & -4.692780 & 0 & 1 & Q & 10.82 & 1.44 & 22.77 & 22.52 & 0.05 & 0.46 & 4.51 & 0.83 & 0.8 \\
255 & 36.687931 & -4.688384 & 0 & 1 & Q & - & - & - & - & - & - & 4.67 & - & - \\\hline\hline\vspace{-2mm}\\
404 & 36.689489 & -4.693379 & 0 & 0 & Q & 10.78 & 1.33 & 22.67 & 22.89 & 0.08 & 0.76 & 4.18 & 0.84 & 8.0 \\
275 & 36.682739 & -4.689313 & 0 & 0 & Q & 10.65 & 1.40 & 23.11 & 23.07 & 0.07 & 0.60 & 4.15 & 0.57 & 0.4
\enddata
\tablecomments{Galaxies are ordered by total integrated magnitude as derived from S{\'e}rsic fits (H$^{tot}_{160}$, see Sec. \ref{sec:photana}) for the observed (top panel), unobserved confirmed (middle panel), and unobserved candidate members (lower panel) respectively. We also show the total integrated magnitudes derived from \textsc{SExtractor} ($H^{auto}_{tot}$) as shown in Fig. \ref{fig:cmd} (see Sec. \ref{sec:photana}). The stellar mass estimates are derived using the total integrated S{\'e}rsic magnitudes (see Sec. \ref{sec:mass}). The 1$\arcsec$ aperture color derived from PSF matched magnitudes ($Y_{105}-H_{160}$) and the major axis $H_{160}$-band sizes ($R^{maj}_{e,H_{160}}$) are given. We show the circularized $B$-band sizes ($R_{e,B}$) and surface brightnesses within $R_{e,B}$ ($\langle I_e \rangle$ in L$_{\odot}$ pc\textsuperscript{-2}) as used for constructing the FP (Sec. \ref{sec:fp}). We also show the axis ratios ($q$) and S{\'e}rsic indices ($n$) as derived from \textsc{galapagos2.2.5b} (see Sec. \ref{sec:photana}). \footnoteref{n14}The IDs and $UVJ$ designations are from \cite{Newman2014}.}  
\end{deluxetable*}

Table \ref{tab:photprops} shows the photometric properties of the 20 spectroscopically confirmed members of JKCS 041, the 16 that we observed with KMOS (top panel), the four confirmed members that we did not observe (middle panel), and the two unobserved candidate members (bottom panel; see Sec. \ref{sec:targ}). The observed ($=1$), and unobserved ($=0$) confirmed ($=1$), and unobserved and unconfirmed ($=0$) galaxies are presented in order of H$^{tot}_{160}$ respectively. The quiescent (Q) and SF designations come from the $UVJ$ diagram in \cite{Newman2014}. We present our derived values of $H^{tot}_{160}$, $R_e$, S{\'e}rsic indices ($n$), and the projected axis ratio $q=\frac{b_e}{a_e}$. Here $a_e$ is the semi-major axis (equivalent to $R^{maj}_e$ as extracted from \textsc{galapagos}), and $b_e$ is the semi-minor axis of the half-light isophote. In this work, we use circularized $R_e$ ($=a_e\sqrt{q}$) values to compare our results for galaxies on the FP with those of previous studies \citep[e.g.][]{Jorgensen2006, Beifiori2017}. As in \cite{Newman2014}, we did not resolve the smallest quiescent galaxy in the sample (ID 255) using our photometric analysis, as its $R_e$ was smaller than one pixel ($< 0\farcs06$).

As the FP is in the rest-$B$ band, and at $z=1.80$ the $H_{160}$ band roughly translates to the rest-$V$ band, we needed to correct the galaxy sizes for the FP. We adopted the prescription derived for the other KCS overdensities in \cite{Chan2016} of $d\log(a_e)/d\log(\lambda) = -0.31 \pm 0.27$. This is consistent with the relation derived by \cite{vanderWel2014} and is roughly constant with redshift. This correction translated to a $\sim 5.6\%$ increase in the galaxy sizes; we plotted this $B$-band circularized $R_e$ ($R_{e,B}$) on the FP. We converted the sizes to kpc assuming $z=1.80$ in our chosen cosmology. We then used $R_{e,B}$ to derive the surface brightness within $R_e$ ($\langle I_e \rangle$) using the $K$-corrected rest-$B$ magnitudes (see Sec. \ref{sec:kcorr}) for the FP.

\subsection{Uncertainties on light-weighted photometric parameters}
\label{sec:photerrs}

To estimate the systematic uncertainties on our derived photometric parameters, we placed simulated galaxies in the \textit{HST} images and extracted them using the method described in Sec. \ref{sec:photana}. We adapted a suite of simulations used in \cite{Chan2016, Chan2017} to run \textsc{galapagos-2.2.5b} on simulated galaxies placed into the $H_{160}$ and $Y_{105}$ images. The simulations worked by creating 2D single S{\'e}rsic profile galaxies from input parameters that represented the variety of objects in the image ($n = 0.2$--$10$, $R_e = 0.012$--$3\arcsec$, $H^{tot}_{160}= 17$--$24$ mag). A noise level was estimated from the image and was added to the simulated galaxy, which was then randomly placed into each band. A mock image was then created for the one source in both bands. Source detection was then done using \textsc{SExtractor}, and light-weighted parameters were derived by running \textsc{galfitm} in the \textsc{galapagos-2.2.5b} software, as for the real galaxies. The input simulated parameters and the output derived parameters were then compared.

Using \textsc{galapagos-2.2.5b}, the uncertainties for the light-weighted parameters derived from the simulations were smaller than those presented in \cite{Newman2014}. All errors quoted below are 1$\sigma$ uncertainties. Directly comparing uncertainties, for $H^{tot}_{160}$ we found an average uncertainty across the whole range (17--24 mag) of $\langle\delta H^{tot}_{160}\rangle = 0.09$ mag. However, we adopted the incremental uncertainties of $\delta H^{tot}_{160} = 0.12$ for $19.5 < H^{tot}_{160} < 21.5$ mag, and  $\delta H^{tot}_{160} = 0.24$ for $21.5 < H^{tot}_{160} < 24$ mag for the galaxies. For $R_e < 0.5\arcsec$, we found uncertainties of $\delta R_e = 7\%$, increasing to $\delta R_e = 13\%$ at $0.5\arcsec < R_e < 1.0\arcsec$, and $\delta R_e = 17\%$ for the most extended profiles at $1.0\arcsec < R_e < 2.0\arcsec$. For $n < 5$, we found $\delta n = 0.2$, and for the profiles with the largest S{\'e}rsic values ($5 < n < 10$) we found uncertainties of $\delta n = 1.0$. We found errors of $\delta q = 0.01$ for all $q$ values. We also found some systematic trends, in that our derived output values of $H^{tot}_{160}$ were slightly fainter than the true value at lower brightnesses, and $n$ and $R_e$ were slightly underestimated at larger values. However, we found that these trends over the range of values covered by the galaxies in JKCS 041 were marginal and well within the quoted $1\sigma$ errors. These uncertainties and trends were similar to those presented in \cite{Chan2016}, which we refer the reader to for a more detailed discussion.

\subsection{$K$-corrections}
\label{sec:kcorr}

To account for redshift and any difference between the emitted and observed spectral regions, we applied $K$-corrections \citep{Hogg1999, Hogg2002} to our derived magnitudes. We split the $K$-correction into two components \citep[as in e.g.,][]{Houghton2012}, such that the total $K$-correction is
\begin{equation} \label{eq:kcorr}
K = K_b + K_c.
\end{equation}
Here the bandpass correction $K_b$ is the reduction of the brightness by ($1 + z$) to account for cosmological expansion. The color correction $K_c$ accounts for the conversion between different rest-frame regions of the spectra, which depends on the underlying stellar population. 

\begin{figure} 
\centering
\includegraphics[width=0.5\textwidth , trim=20 20 20 20,clip]{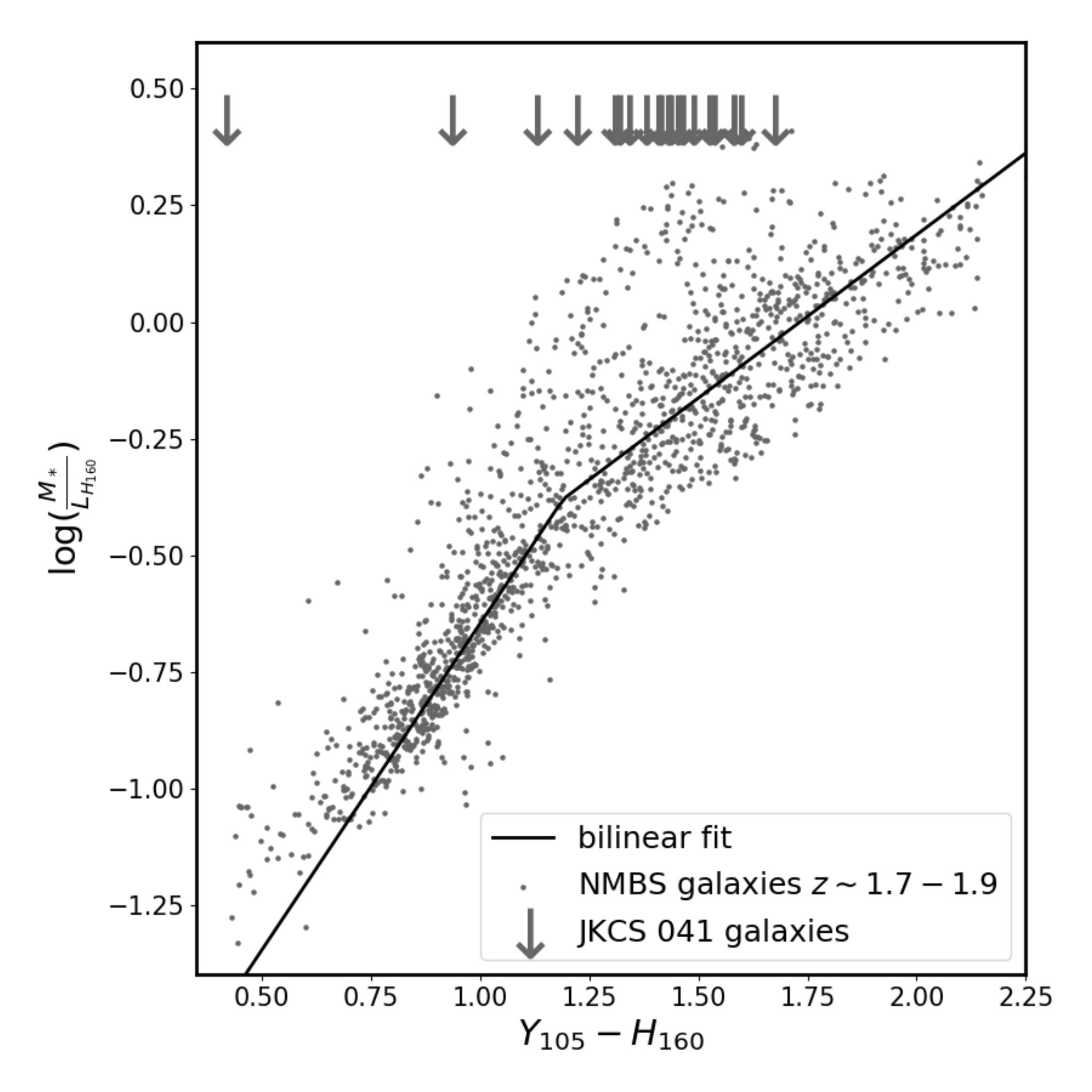}
\caption[]{$M_*/L$-color relation for galaxies from the NEWFIRM Medium Band Survey \citep[NMBS;][]{Whitaker2011} catalogue in the range $1.7 < z < 1.9$. The observed frame colors $Y_{105}-H_{160}$ and $H_{160}$-band luminosities ($L_{H_{160}}$) were derived for the NMBS galaxies as described in \cite{Chan2016, Chan2017}. We used a bilinear fit to define a relation between $\log(M_*/L_{H_{160}})$ and $Y_{105}-H_{160}$, and used this to derive stellar masses for galaxies in JKCS 041 ($Y_{105}-H_{160}$ colors given by gray arrows). See Sec. \ref{sec:mass}.\small}
\label{fig:NMBS}
\end{figure}

To calculate $K_c$ terms for our galaxies, we derived a relation from simple stellar population (SSP) models from which we could convert a galaxy color into a color correction term. We used \cite{Maraston2011} SSP models based on the Medium-resolution Isaac Newton Telescope Library of Empirical Spectra \citep[MILES;][]{Sanchez-Blazquez2006} to derive the relation. We opted for a Chabrier initial mass function \citep[IMF;][]{Chabrier2003}, and used all the possible ages and metallicities of the models ($\sim$ 6 Myrs--15 Gyrs, $\sim$ 0.001--2.5 Z$_{\odot}$ respectively). We then determined magnitudes for each age and metallicity model in the observed $H_{160}$ and $Y_{105}$ bands\footnote{Filter throughput information for \textit{HST}/WFC3 was from \href{ftp://ftp.stsci.edu/cdbs/comp/wfc3/}{ftp://ftp.stsci.edu/cdbs/comp/wfc3/}.}, as observed at $z = 1.80$, using \cite{Hogg2002} Eq. (2). As we wanted to compare our results to those of \cite{Beifiori2017} and use the local Coma FP slopes from \cite{Jorgensen2006}, we opted to correct our galaxy magnitudes to the rest-frame Vega $B$ band \citep{Bessell1990}. We therefore determined the magnitudes of all the models in the $B$ band as observed at $z = 0$ (again  using \citealt{Hogg2002}, Eq. (2)). The $Y_{105} - H_{160}|_{z = 1.80}$ color, and the $K_c$ term ($H_{160}|_{z = 1.80} - B|_{z = 0}$) for all the different age and metallicity models were then linearly fitted to give a relation to convert between the two.

Finally, to determine the $K_c$ terms for each of our galaxies, we calculated colors from the aperture and PSF matched magnitudes derived from both the $Y_{105}$ and $H_{160}$ bands as described at the end of Sec. \ref{sec:photana}. We then determined a corresponding $K_c$ term for our galaxies from their $Y_{105} - H_{160}|_{z = 1.80}$ color and our derived relation. We used the intrinsic scatter of the models in the relation to estimate an error on our $K_c$ terms.

\subsection{Stellar masses}
\label{sec:mass}

To estimate stellar masses for the galaxies in JKCS 041, we used an empirical relation between galaxy color and stellar-mass-to-light ratio ($M_*/L$) as done in \cite{Chan2016, Chan2017}. To derive this relation, we used data from the public catalogue of the NOAO Extremely Wide-Field Infrared Imager (NEWFIRM) Medium Band Survey \citep[NMBS;][]{Whitaker2011}. The NMBS sample has $\sim$ 13,000 galaxies at $z > 1.5$ with accurate rest-frame colors, photometric redshifts (derived with \textsc{eazy}, \citealt{Brammer2008}), and stellar masses derived from spectral energy distribution (SED) fitting of 37 photometric bands using the \textsc{fast} code \citep{Kriek2009}.

From the NMBS catalogue we selected all galaxies in the range $1.7 < z < 1.9$, and a $M_*/L$-color relation was derived in the observer frame to reduce the number of interpolations done to the data \citep{Chan2016, Chan2017}. The derivation was done using \textsc{eazy}; utilizing NMBS photometry and redshifts to fit SEDs,  observed-frame $Y_{105}-H_{160}$ colors, and $H_{160}$-band luminosities ($L_{H_{160}}$). A detailed description of this derivation can be found in \citep{Chan2016}. We show $\log(M_*/L_{H_{160}})$ against the color $Y_{105}-H_{160}$ in Fig. \ref{fig:NMBS}. The $Y_{105}-H_{160}$ color is useful to constrain $M_*/L$ as the bands straddle the 4000$\text{\AA}$ break in the rest frame.

We fitted these galaxies with a bilinear relation. This bilinearity predominately comes from the differences between the red and blue galaxy populations at these redshifts \citep[e.g.][]{Mok2013, Chan2017}. From this relation, we then used our $Y_{105}-H_{160}$ colors for the galaxies, and derived corresponding $\log(M_*/L_{H_{160}})$ values. We then estimated total stellar masses ($\log(M^{tot}_*/M_{\odot})$) using $L_{H_{160}}$ derived from the total integrated S{\'e}rsic magnitudes from \textsc{galapagos2.2.5b} ($H^{tot}_{160}$). These values are shown in Table \ref{tab:photprops}. 

We note that there is significant scatter in the fitted NMBS relation that exists even after a magnitude cut to match the range of values for the JKCS 041 galaxies. Although this increased scatter is expected at higher redshifts, this may mean that our derived light-weighted stellar masses from this relation are uncertain. When deriving errors for our stellar masses we therefore include the scatter on the relation. The uncertainties on our $\log(M^{tot}_*/M_{\odot})$ values are $\delta\log(M^{tot}_*/M_{\odot}) \sim$ 0.12 for the five brightest galaxies (IDs 272, 355, 356, 376, and 657), and $\delta\log(M^{tot}_*/M_{\odot}) \sim$ 0.15 for all others.

To verify our mass estimates, we compared the stellar masses we derived with those from \cite{Newman2014}, that were determined using SED fits to multi-band ground- and space-based images using \textsc{fast} \citep[see][for more details]{Newman2012}. Their masses were scaled to the total magnitude estimated from \textsc{SExtractor} (\texttt{MAG\_AUTO}, see Sec. \ref{sec:photana}). We used these \textsc{SExtractor} total integrated magnitudes from our images ($H^{\rm auto}_{160}$) to derive \texttt{MAG\_AUTO} scaled masses ($\log(M^{\rm auto}_*/M_{\odot})$). Comparing our masses to those of \cite{Newman2014}, we found these to be consistent, differing on average by $\log(M^{\rm auto}_*/M_{\odot}) \sim 0.10$.

%------------------
%SPECTROSCOPIC DATA
%------------------
\section{Spectroscopy}
\label{sec:spec}

\subsection{KMOS observations}
\label{sec:obs}

The overdensity was observed under European Southern Observatory (ESO) programs 095.A-0137(A) and 096.A-0189(A). In total, 16 galaxies were observed over six nights: 17\textsuperscript{th}--19\textsuperscript{th} September 2015 (P95), 10\textsuperscript{th}--12\textsuperscript{th} October 2015 (P96), for $\sim$ 20 hours on source at seeing $< 1 \arcsec$ in the \textit{YJ} band (R $\sim$ 3400, $\lambda \sim$ 1--1.36$\mu$m). To improve the rejection of bad pixels from the final spectra, each exposure was dithered by 0.1--0$\farcs$6. The observed sources comprised 12 quiescent galaxies (one of which was previously an unconfirmed candidate member, ID 772), and four SF galaxies.

\subsection{Data reduction}
\label{sec:specred}

The data reduction of the KCS galaxies utilized a combination of routines from the original KMOS reduction pipeline  \citep[\textsc{spark};][]{Davies2013} and specially written \textsc{python} software \citep{Mendel2015}. This reduction was used for the KCS sample in \cite{Beifiori2017} and will be described further in \cite{Mendel2017}. As an overview, following calibrations and removal of atmospheric absorption using \textsc{molecfit} \citep{Smette2015, Kausch2015}, \textit{HST} images were used to create model source profiles of the galaxies in each IFU. Using the model frames, 1D spectra were optimally extracted \citep{Horne1986} for each galaxy from within $R_e$. When extracting the final spectrum, sigma clipping was performed on all spectra within $R_e$ for all IFUs across the whole 20-hour integration. This produced one spectrum per galaxy for the 16 galaxies that were observed. As well as these individual spectra, there were also 100 bootstrapped realizations of each galaxy spectrum generated from random replacement of all the input spectra (within $R_e$ from the 20-hour observations) prior to sigma clipping and optimal extraction. We used these bootstrapped spectra for quantifying the uncertainties on parameters derived from the spectra.

\subsection{Data analysis}
\label{sec:specana}

\subsubsection{Stellar kinematics}
\label{sec:kinfit}

To derive the stellar kinematics of the galaxies, we fitted the spectra using the Penalized Pixel-Fitting (\textsc{ppxf}\footnote{\href{http://purl.org/cappellari/software}{http://purl.org/cappellari/software}}; \citealt{Cappellari2004}; as upgraded in \citealt{Cappellari2017}) software. \textsc{ppxf} works by fitting an input library of stellar templates or galaxy models to a galaxy spectrum and can be used to derive accurate stellar kinematics from absorptions lines.

The templates chosen to fit the galaxies were from the MILES stellar library, made up of 985 stars, spanning 3525--7500$\text{\AA}$, and covering a large range of stellar atmospheric parameters \citep[][]{Sanchez-Blazquez2006, Falcon-Barroso2011}. We opted to use the high-resolution \citep[2.54$\text{\AA}$ FWHM;][]{Beifiori2011} MILES stellar library rather than SSP models based on the same library, as it has been found that stellar spectral libraries can more accurately recover kinematic parameters than stellar population models when fitting stellar kinematics, as the influence of template mismatch is reduced \citep[][see Sec. 2]{Cappellari2007}. Although in our case, the limiting factor for accurate kinematics is the spectral S/N, and differences between using MILES stellar and MILES-based SSP libraries are negligible, and produce values that are consistent within errors. However, it is important to note that when de-redshifted, our KMOS galaxy spectra increased in resolution from $\sim 3.5 \text{\AA}$ FWHM (as determined from the widths of skylines) to $\sim 1.25 \text{\AA}$ FWHM, corresponding to $\sigma_{\rm KMOS,z=0} \sim 39$ km s$^{-1}$ over a rest-frame wavelength range $\sim$3570--4860$\text{\AA}$ (assuming $z = 1.80$), as compared to $\sigma_{\rm MILES}$ $\sim 77$ km s$^{-1}$. However, due to the stability of the resolution and the wavelength coverage, MILES was still the best template option. We were therefore in the rare case where the galaxy spectral resolution was smaller than the template resolution, and had to account for this when setting up and extracting values from the fit to correctly recover the stellar kinematics.

Prior to fitting with \textsc{ppxf}, we shifted all spectra to the rest frame with initial redshift estimates, then log-rebinned them with the flux rigorously preserved (using \textsc{ppxf} utilities), and the velocity scale set to the minimum of the input spectra (i.e. determined at the reddest pixel, $\sim 39$ km s$^{-1}$). For the template spectra (as $\sigma_{\rm KMOS,z=0} < \sigma_{\rm MILES}/2$), we did not convolve the spectra to match the resolution of the templates but instead had to correct for this difference in resolution after the fit. The templates were clipped to 3000--6000$\text{\AA}$, log-rebinned, and normalized. The templates and galaxy spectra were then given to \textsc{ppxf}, which fitted for velocity, dispersion, and continuum shape simultaneously. The continuum can be fitted with different order additive or multiplicative polynomials. We chose to use additive polynomials in our fits as was done in \cite{Beifiori2017}, and chose a fourth-order polynomial. 

\begin{figure*} 
\centering
\includegraphics[width=0.95\textwidth , trim=20 40 15 35,clip]{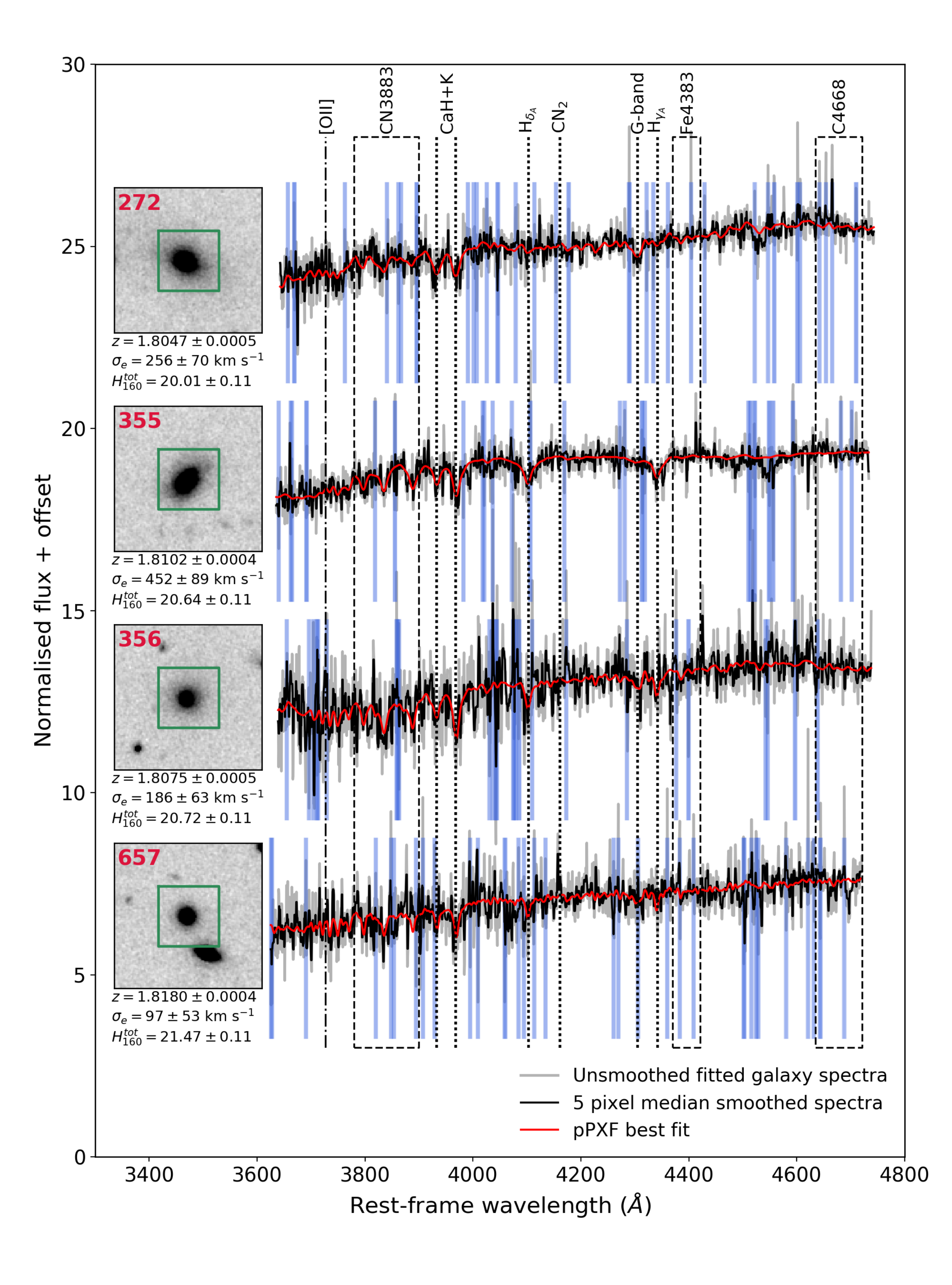}
\caption[]{KMOS spectra of the quiescent galaxies in JKCS 041 for which we derived velocity dispersions. The unsmoothed spectra (grey) were fitted with \textsc{ppxf} (best fit shown in red). We masked pixels (blue bands) based on the subtraction of the five-pixel median smoothed spectra (black) and highlighting $>3\sigma$ outliers from the median of the subtracted spectrum. Next to each spectrum we give the corresponding galaxy ID and $H_{160}$-band postage stamp, along with the redshift and $\sigma_e$ as obtained from the fit. The 1$\sigma$ errors were obtained from the bootstrapped spectra. We plot the spectra from brightest to faintest in $H^{tot}_{160}$, with values derived using \textsc{galapagos} and errors from simulated galaxies (see Secs. \ref{sec:photana} and \ref{sec:photerrs}). Each postage stamp is $6.78\arcsec \times 6.54\arcsec$, the green squares show the size of the KMOS IFUs ($2.8\arcsec \times 2.8\arcsec$). Absorption features (dotted lines) or bands (dashed rectangles), and [OII] emission line for reference (dot-dash line), are indicated. See Sec. \ref{sec:specana}.\small}
\label{fig:qfspec}
\end{figure*}

\renewcommand{\thefigure}{\arabic{figure}}
\addtocounter{figure}{-1}
\begin{figure*} 
\centering
\includegraphics[width=0.95\textwidth , trim=20 40 15 35,clip]{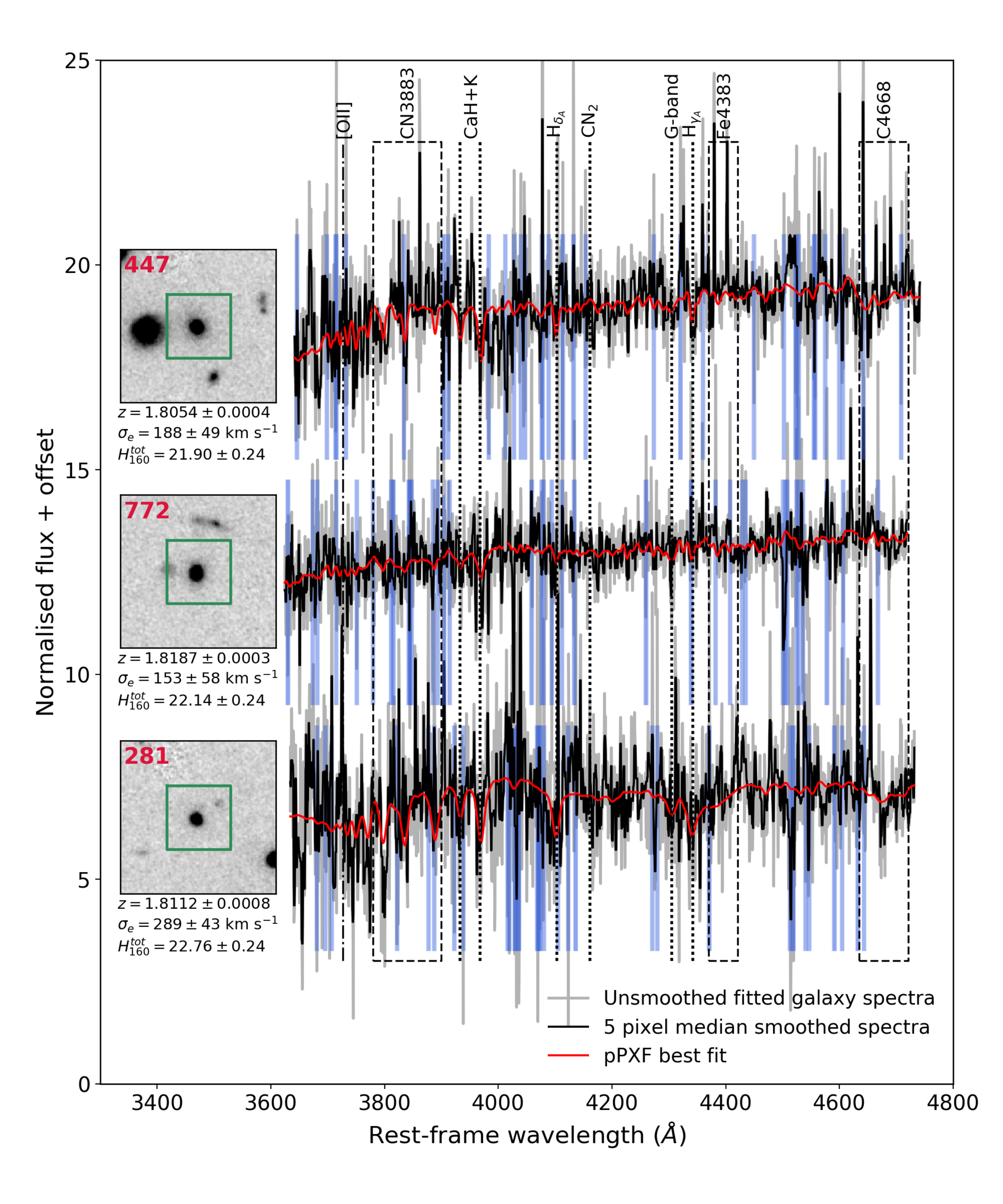}
\caption[]{Continued.\small}
\label{fig:qfspec}
\end{figure*}
\renewcommand{\thefigure}{\arabic{figure}}

\begin{deluxetable*}{cccccccccc}
\tablewidth{0.9\textwidth} 
\tablecaption{\label{tab:spec}Spectroscopic properties of galaxies observed with KMOS.}
\tablehead{\colhead{ID} & \colhead{$z$} & \colhead{$\delta z$} & \colhead{$z$-method} & \colhead{$v_{rel}$ km s$^{-1}$} & \colhead{$\delta v_{rel}$ km s$^{-1}$} & \colhead{$\sigma_e$ km s$^{-1}$} & \colhead{$\delta\sigma_e$ km s$^{-1}$} & \colhead{$\log(M_{dyn}/M_{\odot})$} & \colhead{$\delta\log(M_{dyn}/M_{\odot})$}}
\startdata
272 & 1.8047 & 0.0005 & ppxf & -91 & 24 & 256 & 70 & 11.91 & 0.25 \\
355 & 1.8102 & 0.0004 & ppxf & 142 & 20 & 452 & 89 & 12.05 & 0.18 \\
356 & 1.8075 & 0.0005 & ppxf & 27 & 11 & 186 & 63 & 11.46 & 0.31 \\
657 & 1.8180 & 0.0004 & ppxf & 473 & 20 & 97 & 53 & 10.24 & 0.48 \\
352 & 1.8038 & 0.0027 & ppxf & -130 & 116 & - & - & - & - \\
447 & 1.8054 & 0.0004 & ppxf & -61 & 20 & 188 & 49 & 11.16 & 0.23 \\
286 & 1.8062 & 0.0002 & ppxf & -27 & 11 & - & - & - & - \\
411 & 1.8176 & 0.0012 & ppxf & 456 & 52 & - & - & - & - \\
772 & 1.8187 & 0.0003 & ppxf & 503 & 17 & 153 & 58 & 10.48 & 0.33 \\
387 & 1.8061 & 0.0026 & ppxf & -32 & 111 & - & - & - & - \\
317 & 1.7942 & 0.0034 & ppxf & -543 & 147 & - & - & - & - \\
359 & 1.8057 & 0.0012 & ppxf & -49 & 52 & - & - & - & - \\
693 & 1.8239 & 0.0018 & ppxf & 722 & 76 & - & - & - & - \\
281 & 1.8112 & 0.0008 & ppxf & 185 & 35 & 289 & 43 & 11.00 & 0.13 \\
531 & 1.8157 & 0.0006 & emission & 376 & 27 & - & - & - & - \\
332 & 1.8001 & 0.0063 & emission & -289 & 270 & - & - & - & -
\enddata
\tablecomments{The redshift and method of its determination are presented, along with the velocity relative to the median redshift ($v_{rel}$). The seven reliable $\sigma_e$ values and their corresponding dynamical masses (derived using $B$-band sizes) are also given. See Sec. \ref{sec:specana}.}
\end{deluxetable*}

Some areas of the spectra were affected by strong sky line residuals even after reduction. The three worst affected bands, when shifted to the rest frame (assuming $z = 1.80$), corresponded roughly to $\sim$ 3570--3750, 4040--4130, and 4460--4600$\text{\AA}$. To determine the bad pixels to mask from the kinematic fit, we first median smoothed each spectrum by five pixels and subtracted the smoothed spectrum from its unsmoothed counterpart. Using this residual subtracted spectrum, we then selected all those pixels that deviated from the mean by $>3\sigma$. We then `grew' these selected bad pixels by $\pm$ one pixel, creating small masks around them. Masking these bad pixel regions, we then fitted the unsmoothed spectra with \textsc{ppxf}. With this method, we masked between 5--7$\%$ of pixels in each spectrum. We found this method could reliably mask bad pixels or poor regions of the spectra that were heavily affected by sky lines, and enabled us to improve the fits of the galaxies. We performed various tests to check the reliability of this masking method. We tried different combinations of three-, five-, and seven-pixel median smoothing, and selected 2- and 3$\sigma$ outliers from the mean of the subtracted spectrum, but found all methods produced comparable results.

Using this preparation of the templates and spectra, and using the masking method described above, we fitted each galaxy and its 100 bootstrapped spectra with \textsc{ppxf} to measure the kinematics. To estimate errors on the kinematic fits, we fitted the 100 bootstrapped realizations of the galaxy spectra. To account for any systematics introduced by our choice of order polynomial, we fitted the bootstraps with random order additive polynomials in the range two to eight. The quoted errors, as determined from the median-absolute-deviation (MAD) of the results from the bootstrapped spectra and converted to 1$\sigma$ errors assuming a normal distribution (1$\sigma=1.4826\times$MAD), reflected these systematic uncertainties. 

The errors on the kinematic measurements we derived reflect the ability to reproduce consistent results from the many bootstrapped realizations of each galaxy spectrum. Tests performed for the other KCS targets showed that some of the best quality spectra were extracted from more compact objects and not only the brightest galaxies. For larger galaxies with more spectra within $R_e$, the bootstrapped spectra vary more, sometimes producing larger errors despite their higher S/N. Nonetheless, this remains the best estimation of systematic errors for the spectra in our sample. We deemed a stellar velocity dispersion reliable if we were able to fit $> 70\%$ of the bootstrapped spectra and the relative error (i.e. variation in values) we derived from the bootstrapped spectra was $\frac{\delta \sigma_e}{\sigma_e} \lesssim 50\%$.

To account for the differing resolutions of the templates and data, the stellar velocity dispersions as measured within $R_e$ for the galaxies ($\sigma_e$) had to be determined from the observed value of the dispersion ($\sigma_{obs}$ -- measured by \textsc{ppxf} in the case of no convolution). This was done using
\begin{equation} \label{eq:csig}
\sigma_e = \sqrt{\sigma_{obs}^2 - \sigma_{diff}^2}.
\end{equation}
Where $\sigma_{diff}$ is given by
\begin{equation} \label{eq:dsig}
\sigma_{diff}^2 = \sigma_{\rm KMOS,z=0}^2 - \sigma_{\rm MILES}^2.
\end{equation}
Following these corrections, and accounting for the errors derived from the distribution of values from the bootstrapped spectra, we derived reliable velocity dispersions for seven of the galaxies.

\subsubsection{KMOS redshifts}
\label{sec:z}

With the KMOS spectra, we improved on the accuracy of the grism redshift measurements for all the 16 observed galaxies. For most (mainly quiescent) galaxies, this was done using kinematic fits. Even for those galaxies for which we could not derive a reliable value of $\sigma_e$, in most cases we were able to determine improved redshift measurements. Velocity measurements were taken from the fits of the individual galaxies and used to determine redshifts using \cite{Cappellari2009} Eq. (2). From kinematic fitting, we made improvements by around a factor of $\sim$4--5 for most compared to the grism redshifts. For two SF galaxies (IDs 332, 531), we used the strong [OII] $\lambda 3726-3729 \text{\AA}$ doublet to derive a redshift. All redshift measurements determined for the observed spectra are in Table \ref{tab:spec} along with the method used to derive them.

\subsubsection{Properties of the spectra}
\label{sec:specprops}

We present the spectra of the seven quiescent galaxies for which we have reliable stellar velocity dispersions in Fig. \ref{fig:qfspec}. We show the $H_{160}$ image postage stamp of each galaxy with its corresponding ID, and present the spectra in order of $H^{tot}_{160}$ S{\'e}rsic magnitude. Each postage stamp has size $6.78\arcsec \times 6.54\arcsec$ and we have overlaid the size of the KMOS IFUs for reference (green squares, $2.8\arcsec \times 2.8\arcsec$). We show both unsmoothed spectra (grey; that was used for fitting) and five-pixel median smoothed spectra (black). We also indicate the pixels masked from the fit (blue bands; see Sec. \ref{sec:kinfit}). We show the \textsc{ppxf} fits to the spectra in red, and indicate absorption features (dotted lines) or bands (dashed rectangles), and [OII] emission line for reference (dot-dash line). For each galaxy, we give its corresponding redshift and $\sigma_e$ as determined from the fit, with 1$\sigma$ errors from the bootstrap spectra. As we show normalized flux, we have included the $H^{tot}_{160}$ mag for reference.

Table \ref{tab:spec} gives the spectroscopic properties of all the 16 observed galaxies in JKCS 041. The galaxies are ordered from brightest to faintest in $H^{tot}_{160}$. The redshift values are given along with the method used to derived them (see Sec. \ref{sec:z}), as are the measurements for the seven galaxies from which we derived a stellar velocity dispersion.

\subsubsection{Dynamical masses}
\label{sec:mdyn}

For the seven galaxies with reliable $\sigma_e$ measurements, we estimated dynamical masses using the virial relation of \cite{Cappellari2006} \citep[as done in][]{Beifiori2014, Beifiori2017}:
\begin{equation} \label{eq:Mdyn}
M_{dyn} = \frac{\beta (n) \sigma_e^2 R_e}{G},
\end{equation}
where $\beta(n)$ is dependent on the S{\'e}rsic index \citep{Bertin2002a}. However, this approach does have limitations, the effect of the dark matter contribution is unconstrained and the $\beta(n)$ is derived from idealized, isotropic and spherical galaxies. As we used a rest-$B$ band FP and used these dynamical masses for investigating the effect of structural evolution of the FP shift (see Sec. \ref{sec:massplane} and \ref{sec:evoage}), we used a $B$-band $R_e$ (see Sec. \ref{sec:fp}) to derive our dynamical masses. The dynamical mass estimates and errors for the seven galaxies are shown in Table \ref{tab:spec}. 

%-----------------
%Fundamental Plane
%-----------------
\section{Fundamental plane}
\label{sec:fp}

\begin{figure*} 
\centering
\includegraphics[width=0.5\textwidth , trim=20 20 20 20,clip]{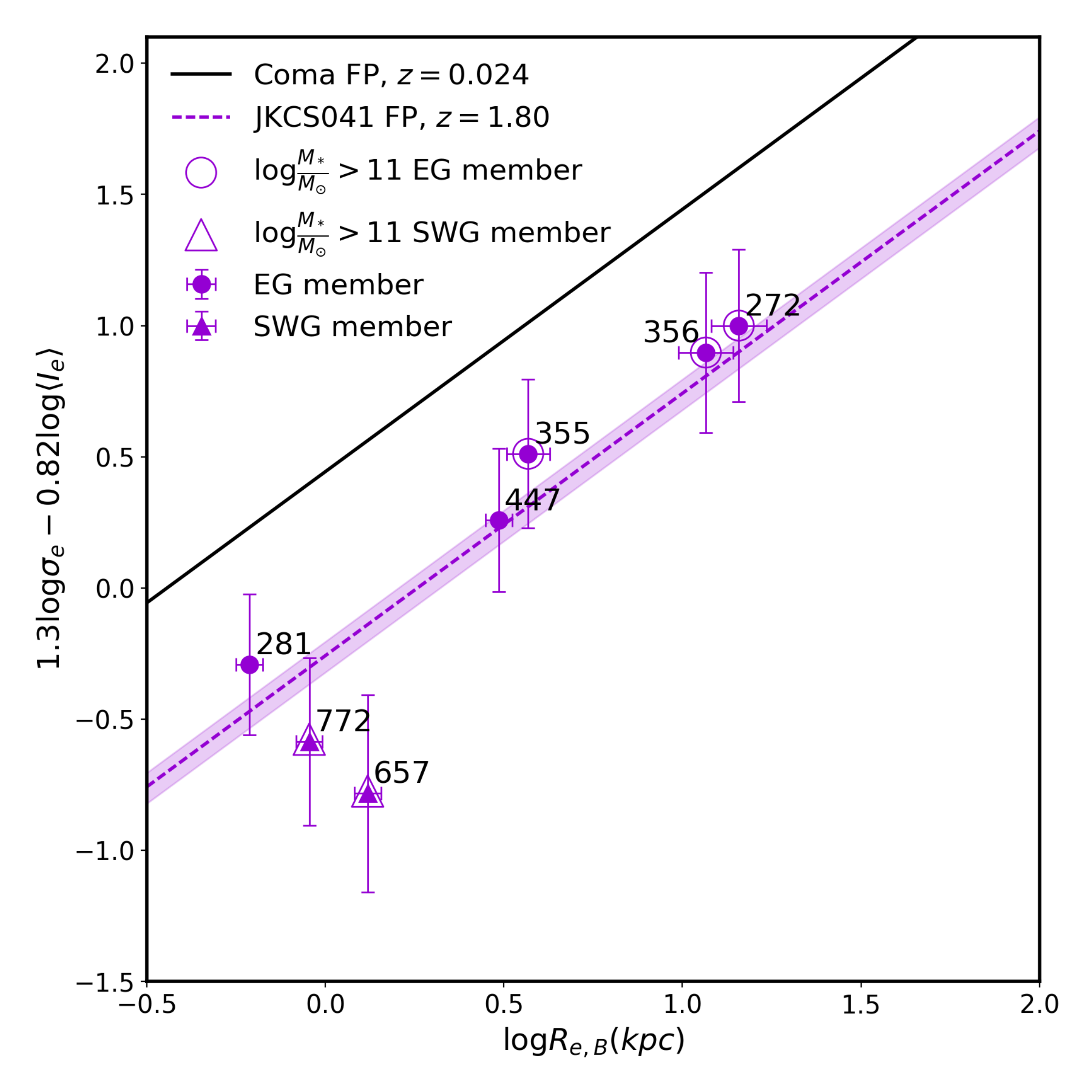}
\caption[]{FP for the seven quiescent galaxies in JKCS 041 (in rest-$B$ band) for which we have derived stellar velocity dispersion measurements. The galaxies in different regions of the overdensity (as determined in 2D from Fig. \ref{fig:hst}) are separated into a group extending eastward (EG; circles) and toward the south west (SWG; triangles; see Sec. \ref{sec:struct}). The best fit (dashed line) and 1$\sigma$ error on the measurement of the FP zero point from bootstrapping (shaded region) are shown. The local fit for Coma from \cite{Jorgensen2006} is shown for comparison; we have adopted the same slopes for JKCS 041. We also show the high light-weighted stellar mass ($\log(M^{tot}_*/M_{\odot}) >$ 11; large unfilled symbols) galaxies in the sample. See Sec. \ref{sec:fp}.\small}
\label{fig:fp}
\end{figure*}

\begin{figure*}
\includegraphics[width=0.49\textwidth , trim=20 20 20 20 ,clip]{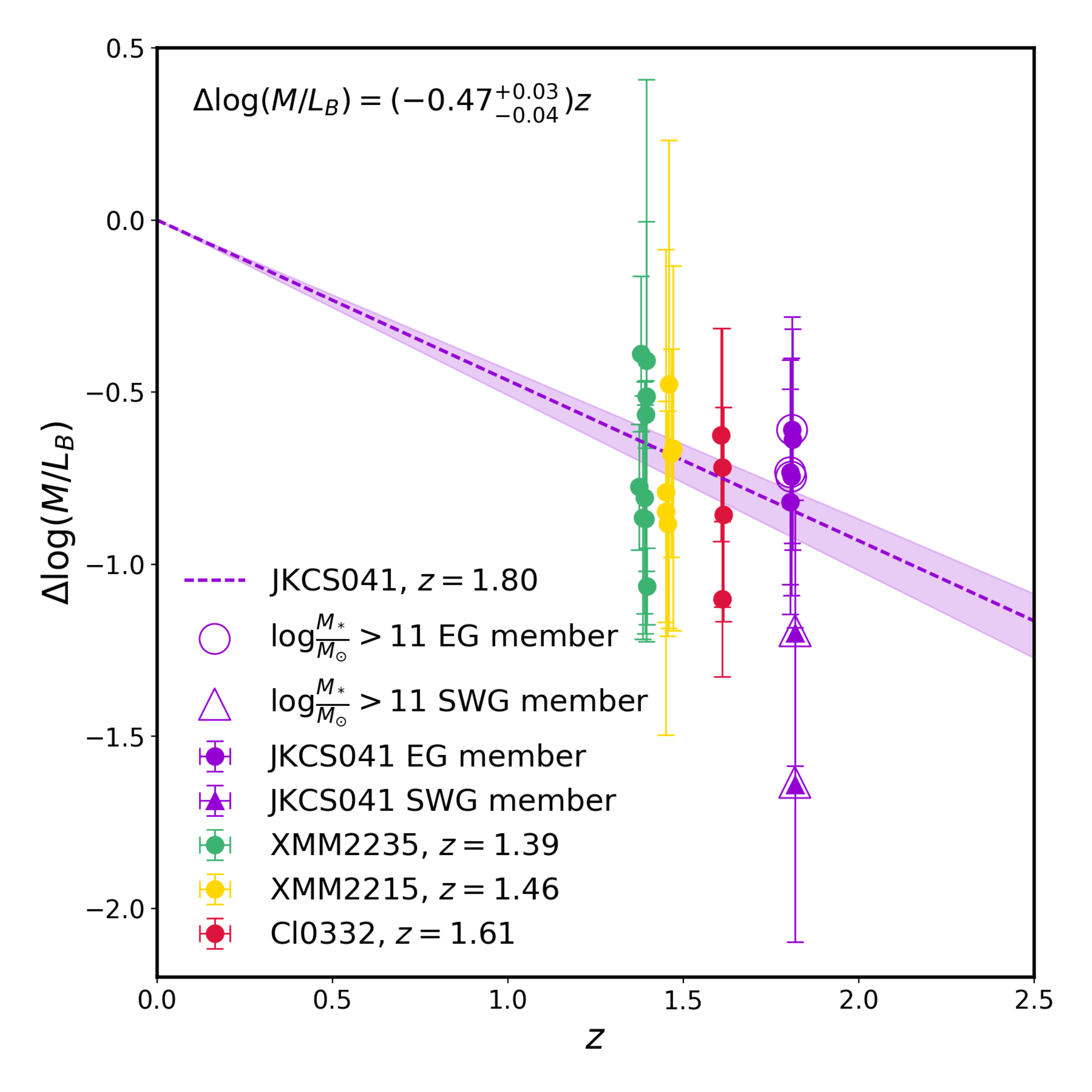}
\includegraphics[width=0.49\textwidth , trim=20 20 20 20,clip]{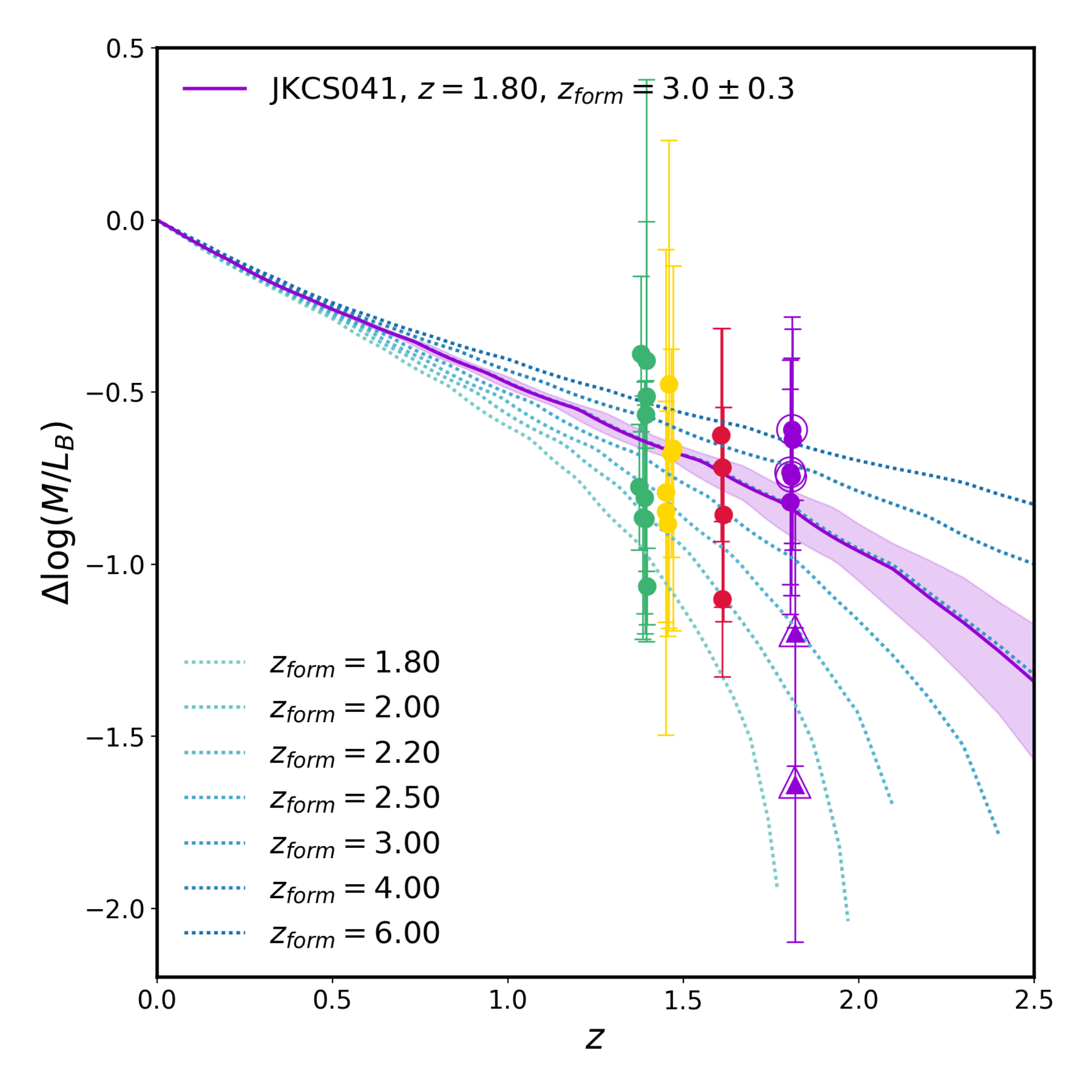}	
\caption[]{$M/L$ evolution (rest-frame $B$-band) derived from the FP (Eqs. \ref{eq:MtL} $\&$ \ref{eq:cz}) as a function of redshift for seven galaxies in JKCS 041 (violet, symbols as in Fig. \ref{fig:fp}). Galaxies in three other KCS overdensities are shown for comparison \citep{Beifiori2017}. \textit{Left:} The best fit (dashed line) and 1$\sigma$ errors from bootstrapping (shaded region) are shown. \textit{Right:} Interpolating SSP models \citep{Maraston2005} using \textsc{EzGal} \citep{Mancone2012}, $M/L$ evolutionary tracks were built up as a function of redshift for different $z_{form}$ values (see Sec. \ref{sec:fp}). Overplotting these tracks (dotted lines), and determining the best fitting $z_{form}$ (solid line) and 1$\sigma$ errors from bootstraps (shaded region) based on the track intersection with the best fit from the left panel, we derived a mean $z_{form} = 3.0\pm0.3$ ($1.4\pm0.2$ Gyrs mean age) for these seven galaxies in JKCS 041.\small}
\label{fig:MtL}
\end{figure*}

The FP for our seven galaxies with reliable $\sigma_e$ measurements is shown in Figure \ref{fig:fp}. We used the following form of the FP so as to compare our results with the local FP of \cite{Jorgensen2006} and the KCS FP at $1.39 < z < 1.61$ from \cite{Beifiori2017}:
\begin{equation} \label{eq:fp}
\log R_e = a\log \sigma_e + b\log \langle I_e \rangle + c_z.
\end{equation}
Here $R_e$ is the circularized effective radius (kpc), $\sigma_e$ is the velocity dispersion within $R_e$ (km s\textsuperscript{-1}), $\langle I_e \rangle$ is average surface brightness within $R_e$ (L$_{\odot}$ pc\textsuperscript{-2}), and $c_z$ is the redshift dependent zero-point.

We fixed the slopes of the FP to the coefficients found for the Coma cluster, at $z=0.024$ in the local-$B$ band, of $a = 1.30\pm0.08$ and $b = -0.82\pm0.03$ \citep{Jorgensen2006}. We opted to use fixed slopes of the FP, mainly as we had too few points to constrain any potential tilt. However, the assumption of a non-tilting FP, at least to $z \sim 1$, is supported by various studies \citep{Wuyts2004, Holden2010, Bezanson2015, Oldham2017}. The zero-point of the FP was fitted using a least-squares method\footnote{Using the orthogonal distance regression package in \textsc{scipy} in \textsc{python}.} that accounted for the errors in the values on both axes (dashed line). We obtained errors via a bootstrapping method \citep{Jorgensen1996}, selecting random samples of the points with replacement and determining their best fit. The zero-point we obtained from the fit to the FP was $c_z = -0.26\substack{+0.05\\-0.07}$, with 1$\sigma$ uncertainties from bootstrapping the fit (shaded region). 

Using the zero-point shift of the FP, the evolution in $M/L$ $(\Delta \log (M/L_B))$ as a function of redshift can be investigated \citep[e.g.,][]{vanDokkum1996}. This approach makes the assumptions that the ETGs are homologous, the evolution of $c_z$ only depends on changes in $M/L$, and that $a$ and $b$ are redshift independent. The evolution in $M/L$ is expressed as follows:
\begin{equation} \label{eq:MtL}
\begin{aligned}
\Delta \log(M/L_B) = & \log(M/L)_z - \log(M/L)_{Coma} \\
= & (c_z - c_{Coma})/b.
\end{aligned}
\end{equation}
Where $c_{Coma}$ is that defined in \cite{Jorgensen2006}, and $c_z$ values are calculated for each galaxy using the following equation \citep[see e.g.,][]{vanDokkum2007},
\begin{equation} \label{eq:cz}
c_z = \log R_e - (a\log \sigma_e + b\log \langle I_e \rangle).
\end{equation}
This conversion gives a $\Delta \log(M/L_B)$ value for each of the seven galaxies on the FP that is plotted as a function of their redshifts (Fig. \ref{fig:MtL}). Again, we fitted the points using a least-squares method accounting for errors on both axes (dashed line) and derived 1$\sigma$ errors from bootstrapping (shaded region). We found $\Delta \log(M/L_B) = (-0.47\substack{+0.03\\-0.04}) z$ for the seven galaxies in JKCS 041 on the FP (median $z = 1.808$).

\subsection{Derivation of FP ages}
\label{sec:ages}

To determine a formation epoch, and thus age of the galaxies on the FP, we interpolated SSP models at different formation redshifts ($z_{form}$) to get $M/L$ values. We used \textsc{EzGal} \citep{Mancone2012} to interpolate solar metallicity SSP models with a \cite{Salpeter1955} IMF from \cite{Maraston2005} to derive model $M/L$ values in the $B$ band at different formation epochs. We found the age results for JKCS 041 were consistent within errors when using different models (\citealt{Conroy2009, Conroy2010, Bruzual2003}) and IMFs (Salpeter, Chabrier, \citealt{Kroupa2001}). As we use the relative $M/L$, the effects of using different IMFs is expected to be negligible. We refer the reader to Appendix C of \cite{Beifiori2017} for further tests with different SSP models and metallicities, and discussion on how this assumption only affects the ages derived within the stated errors.

We show tracks of $\Delta \log(M/L_B)$ as a function of redshift for different $z_{form}$ values calculated with \textsc{EzGal} in Fig. \ref{fig:MtL}. From interpolating the models for many $z_{form}$ values, we derived the mean best fitting formation epoch for our galaxies as determined from the intersection of the best fitting line in the left panel, and the corresponding model track (only a few of which are shown) in the right panel. The best fitting formation redshift for the seven galaxies was $z_{form} = 3.0\pm0.3$, with errors determined from the model tracks intersecting the 1$\sigma$ uncertainties of the fit from bootstrapping. This $z_{form}$ value corresponds to a mean age for the seven galaxies in JKCS 041 on the FP of $1.4\pm0.2$ Gyrs.

In order to compare JKCS 041 age estimates to red-sequence galaxy ages derived in the three other KCS overdensities in Sec. \ref{sec:compkcs}, we also show the $\Delta(\log M/L_B)$-$z$ values for XMM2235 ($z = 1.39$, green),  XMM2215 ($z = 1.46$, yellow), and Cl0332 \citep[$z = 1.6$, red;][]{Beifiori2017} in Fig. \ref{fig:MtL}. We investigated the FP ages for the most massive galaxies ($\log (M^{tot}_*/M_{\odot}) > 11$; larger unfilled symbols in Figs. \ref{fig:fp} and \ref{fig:MtL}) in JKCS 041 to compare our results, and found comparable values for the five most massive galaxies (as derived from stellar light) to those we obtained for all seven galaxies ($z_{form} = 2.8\substack{+0.5 \\ -0.4}$, mean age $1.2\pm 0.4$ Gyrs).

\subsection{Structural evolutionary effects on the FP zero-point evolution}
\label{sec:zpevo}

\subsubsection{FP zero-point and luminosity evolution}
\label{sec:zpLevo}

To understand the effects of galaxy structural evolution on the change in zero-point of the FP, we used the method described in \cite{Saglia2010, Saglia2016, Beifiori2017}. The FP zero-point can be derived from the change with redshift of the structural evolutionary terms of $R_e$, $\sigma_e$, and luminosity ($L_{\rm FP,SE}$, where SE is structural evolution). The variation of the luminosity from this structural evolution can be expressed as
\begin{equation} \label{eq:Lstrevo}
\begin{aligned}
\Delta \log L_{\rm FP,SE} = &\left( \frac{2b+1}{b}\nu - \frac{a}{b}\mu - \frac{1}{b} \eta\prime \right) \log(1+z)\\
= & \chi\log(1+z).
\end{aligned}
\end{equation}
Here $a$ and $b$ are the coefficients from the FP (Eq. \ref{eq:fp}), $\nu$ and $\mu$ are the slopes of the evolution of sizes and velocity dispersions with redshift, $\eta\prime$ is related to the slope of the $\log(M/L)$ evolution with $\log(1+z)$ by $\eta\prime = \eta \times b$, and $\chi=\left( \frac{2b+1}{b}\nu - \frac{a}{b}\mu - \frac{1}{b} \eta\prime \right)$. We refer the reader to \cite{Beifiori2017} for more details on this derivation.

\begin{figure*} 
	\includegraphics[width=0.49\textwidth , trim=0 0 0 0 ,clip]{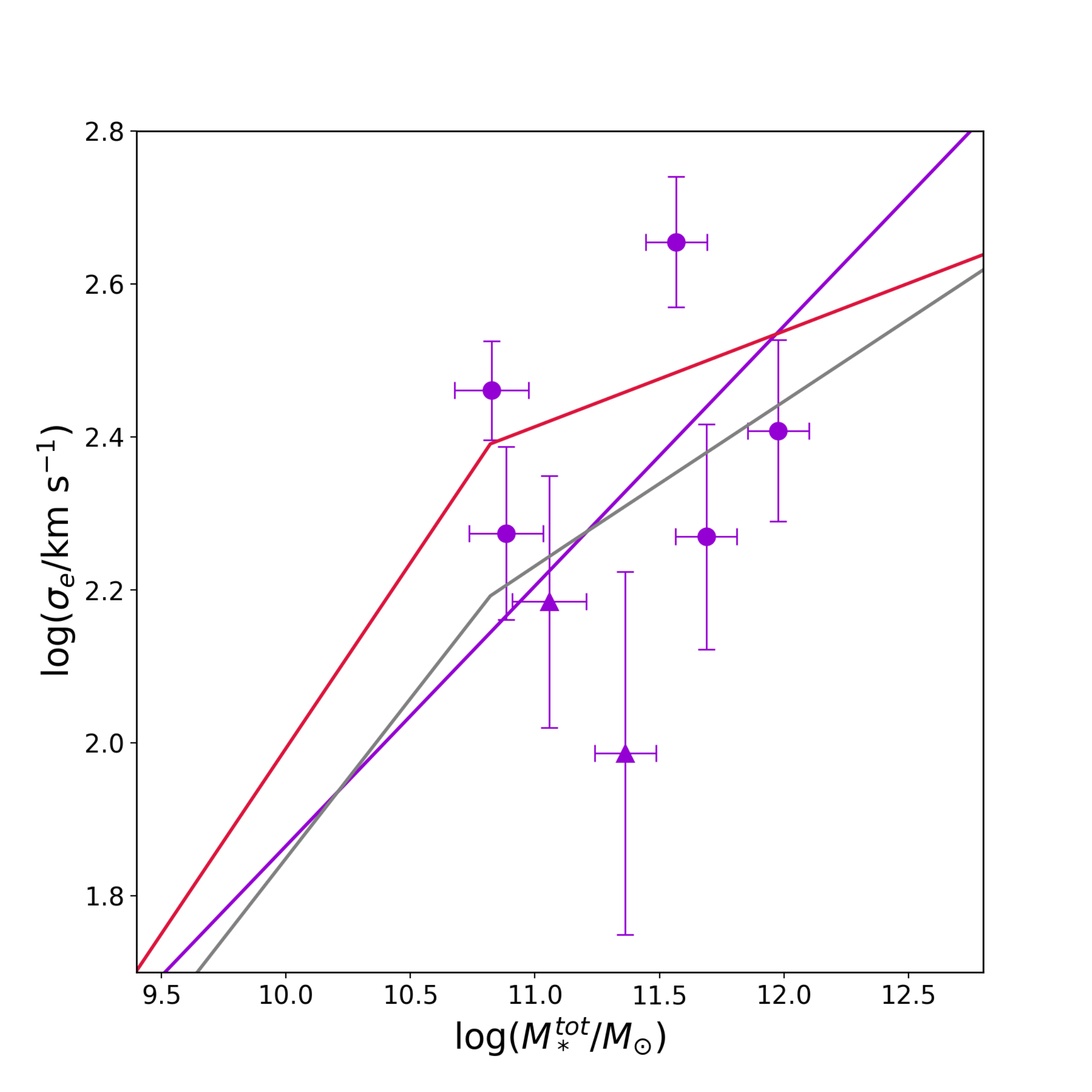}
	\includegraphics[width=0.49\textwidth , trim=0 0 0 0 ,clip]{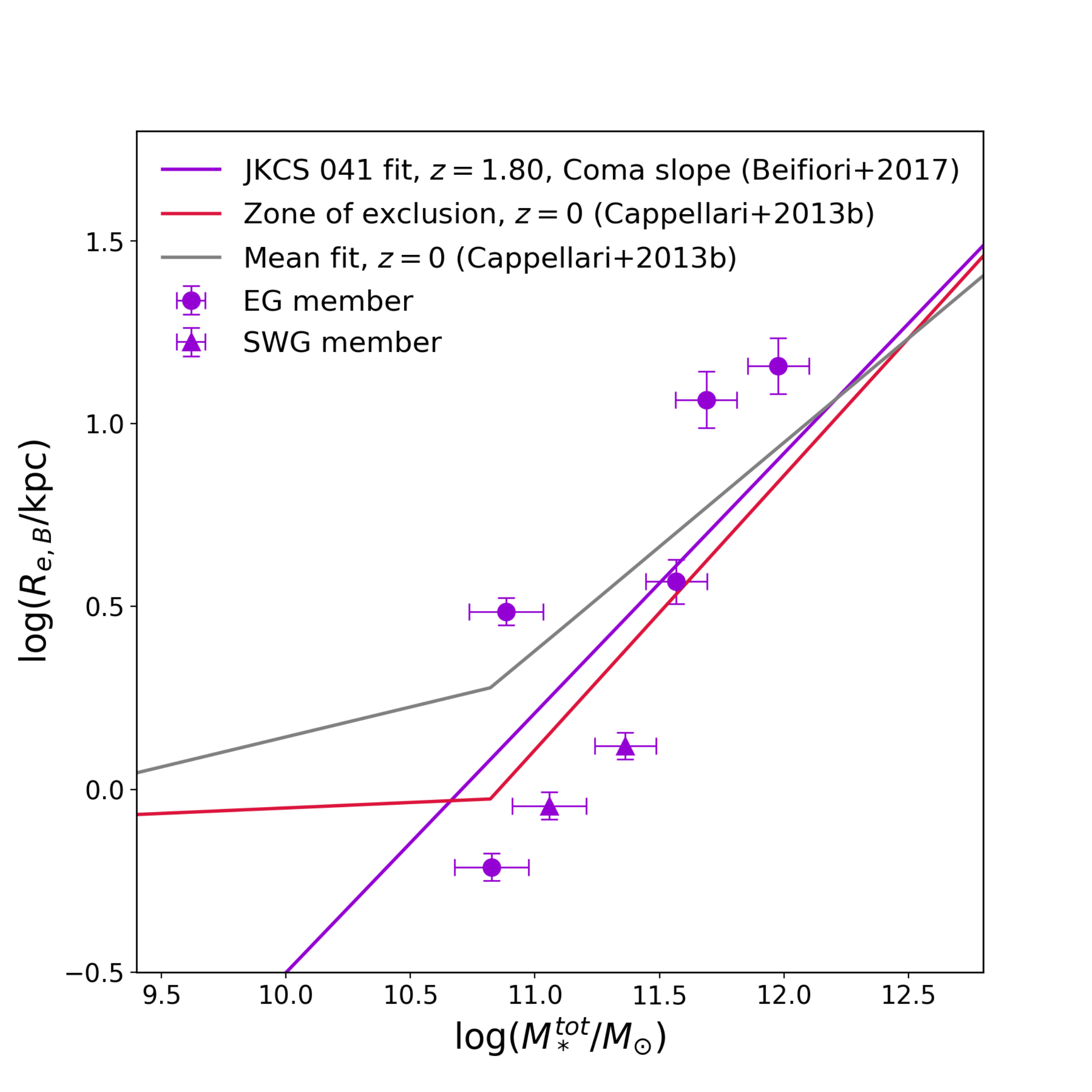}\\
	\includegraphics[width=0.49\textwidth , trim=0 0 0 0 ,clip]{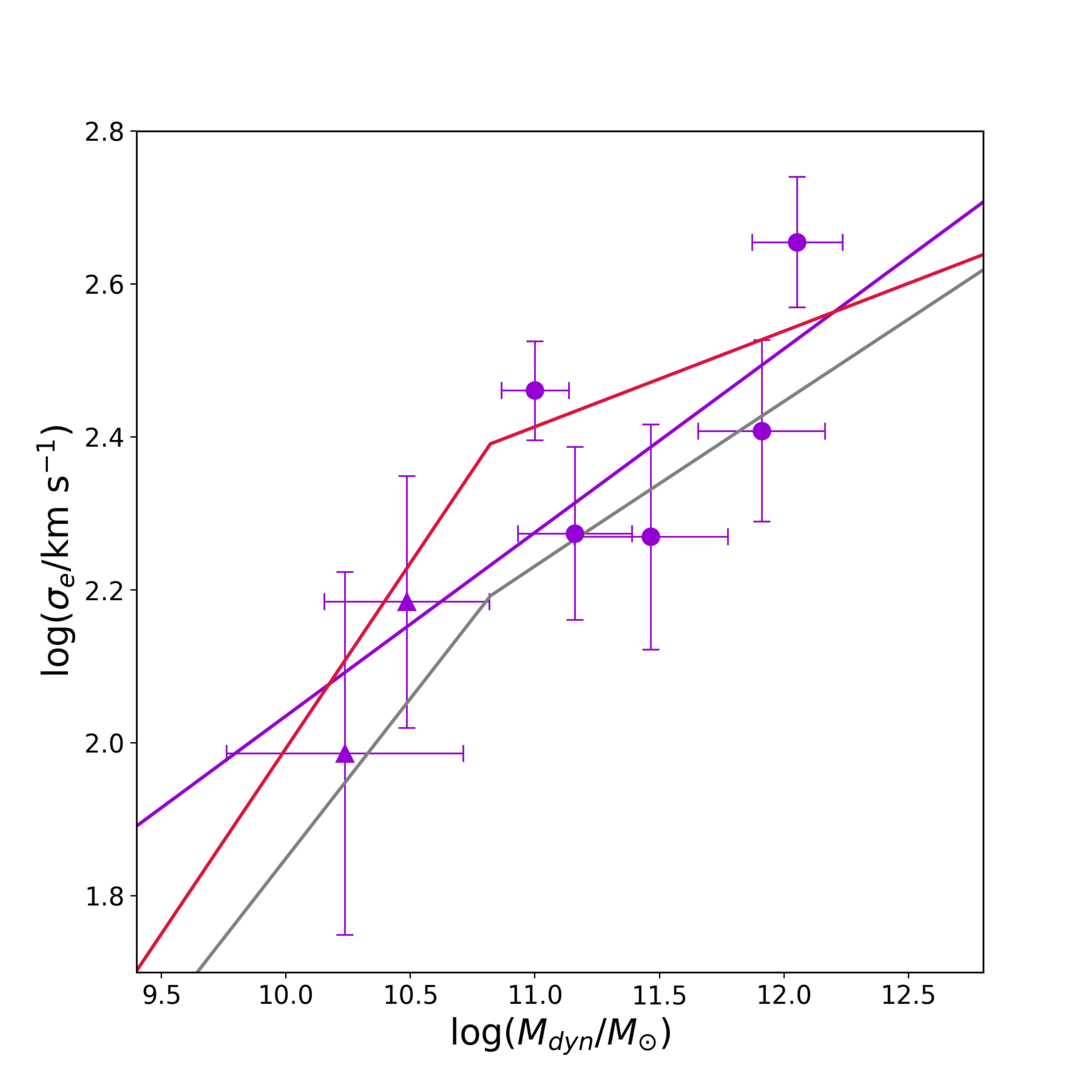}
	\includegraphics[width=0.49\textwidth , trim=0 0 0 0,clip]{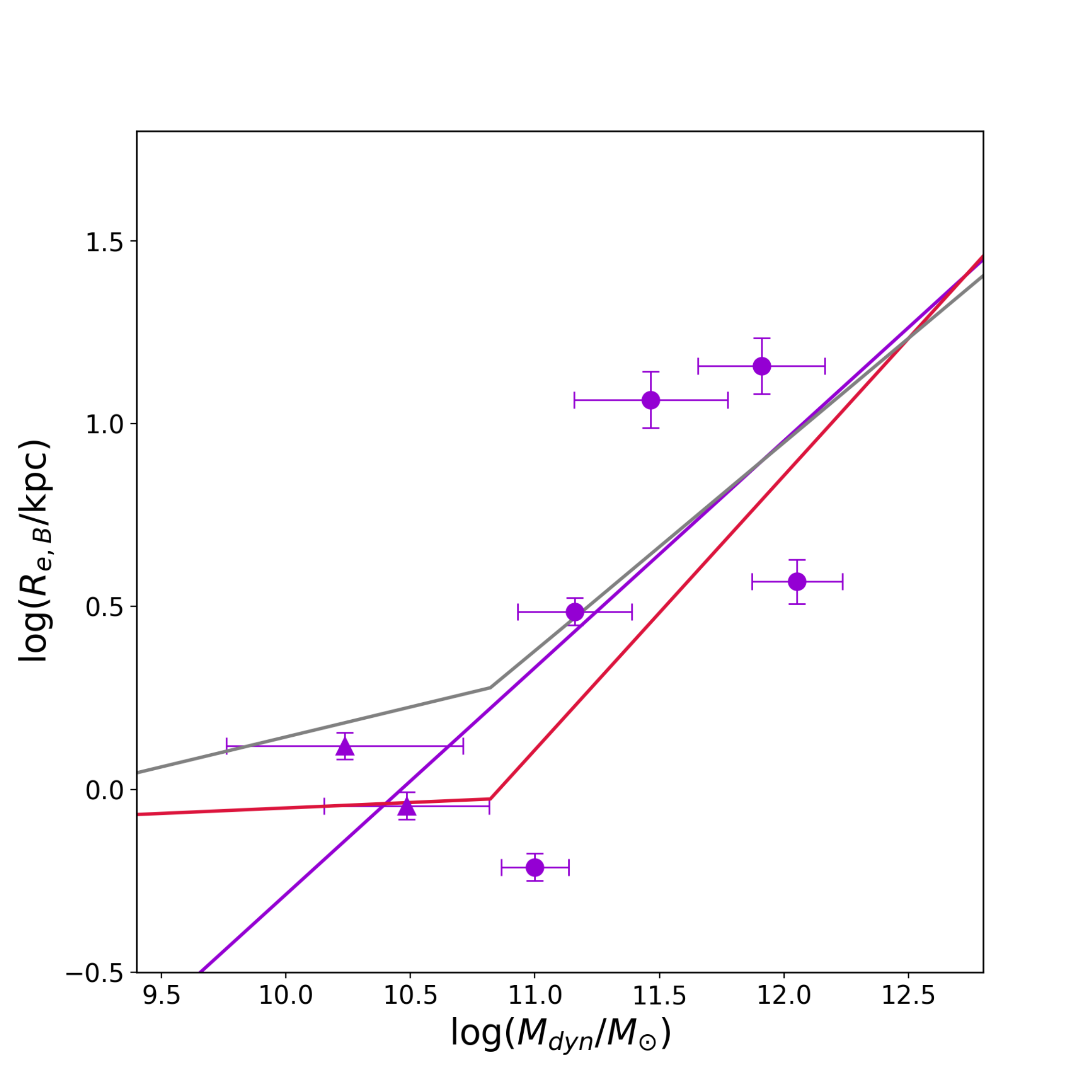}\\
\caption[]{Mass-velocity dispersion ($M$-$\sigma_e$) and mass-size ($M$-$R_e$) relations for the seven galaxies in JKCS 041 with reliable $\sigma_e$ measurements in the EG (circles) or SWG (triangles). \textit{Top:} Relations derived using the $M^{tot}_*$ mass estimate from total integrated magnitude from S{\'e}rsic fits (see Sec. \ref{sec:mass}). \textit{Bottom:} $M$-$\sigma_e$ and $M$-$R_e$ relations derived using dynamical masses ($M_{dyn}$; see Sec. \ref{sec:mdyn}). For reference, we show zone of exclusion (red line) and mean fit to local galaxies (gray line) from \cite{Cappellari2013b}. We linearly fit the galaxies to the local slopes of Coma as derived in \cite{Beifiori2017} (violet lines). See Sec. \ref{sec:zpevo}.\small}
\label{fig:MsigMRe}
\end{figure*}

\begin{deluxetable*}{ccc}
\tablewidth{0.6\textwidth} 
\tablecaption{\label{tab:evo}Slopes derived for FP luminosity zero-point and structural evolutionary parameters.}	
\tablehead{\colhead{Relation} & \multicolumn{2}{c}{Slope}}\\
								& \colhead{$M_{dyn}$} & \colhead{$M_*^{tot}$}
\startdata
 &$M_{dyn}$ & $M_*^{tot}$\\\hline \vspace{-2mm}\\
$\Delta \log L_{\rm FP}=-\eta \log(1+z)$ & \multicolumn{2}{c}{1.88 $\substack{+0.13\\-0.17}$} \\
$\Delta \log R_{e,\rm MN} \propto \nu \log(1+z)$&$-0.61\substack{+0.41\\-0.20}$&$-1.00\substack{+0.20\\-0.19}$\\
$\Delta \log \sigma_{e,\rm MN} \propto \mu \log(1+z)$&$0.26\substack{+0.07\\-0.13}$&$-0.01\substack{+0.08\\-0.14}$\\
$\Delta \log L_{\rm FP,SE} = \chi\log(1+z)$&$1.82 \pm 0.42$&$1.08 \pm 0.33$
\enddata
\tablecomments{We show the slopes for the change in FP luminosity without accounting for structural evolution of the galaxies ($\Delta \log L_{\rm FP}$) and with ($\Delta \log L_{\rm FP,SE}$). We also show the slopes derived for the change in the mass-normalized $R_e$ ($\nu$) and $\sigma_e$ ($\mu$) values with redshift for both the $M_{dyn}$- and $M^{tot}_*$-normalizations, used to determine $\Delta \log L_{\rm FP,SE}$. See Sec. \ref{sec:zpevo}.} %INCLUDED
\end{deluxetable*}

To test the contribution of each structural evolution parameter to the FP zero-point, we first determined the value of just the change in luminosity due to the evolving stellar population. To do this, we assumed that the changes in $\Delta \log R_e$ and $\Delta \log \sigma_e$ were zero, leaving $\Delta \log L_{\rm FP}=-\frac{1}{b}\eta \prime \log(1+z)=-\eta \log(1+z)$. The slope we derived ($\eta = 1.88\substack{+0.13 \\ -0.17}$) is given in Table \ref{tab:evo}.

\subsubsection{Mass-size and mass-sigma relations for JKCS 041}
\label{sec:massplane}

We can use the mass-velocity dispersion ($M$-$\sigma_e$) and mass-size ($M$-$R_e$) relations to study the effects of structural evolution with redshift, as compared to Coma, for the seven galaxies in JKCS 041 with reliable $\sigma_e$ measurements \citep[following the method of][]{Saglia2010, Saglia2016, Beifiori2014, Beifiori2017}. The mass plane (MP) is the narrow relation between mass, $\sigma_e$, and $R_e$, and it follows the scalar viral relation $M \propto \sigma^2R_e$ \citep{Cappellari2013a}. The FP for galaxies in the local Universe was found to be due to the virial relation and a smooth variation of galaxy properties \citep{Cappellari2013a}. As a result, useful information about the properties of galaxies comes from inhomogeneities about the plane, and specifically non-edge-on projections \citep[e.g.,][]{Cappellari2013b, Cappellari2016, Beifiori2017}. We show two such projections of the MP in Fig. \ref{fig:MsigMRe}, the $M$-$\sigma_e$ and $M$-$R_e$ relations, with circularized $R_{e,B}$ values as used for the FP. The relations for both the stellar masses derived from our total integrated S{\'e}rsic magnitudes ($M_*^{tot}$-$\sigma_{e}$ and $M_*^{tot}$-$R_{e,B}$; see Sec. \ref{sec:mass}), and the dynamical masses derived from the $\sigma_e$ values ($M_{dyn}$-$\sigma_e$ and $M_{dyn}$-$R_{e,B}$; see Sec. \ref{sec:mdyn}) are shown.

For reference, we show the zone of exclusion for local galaxies, where local ETGs are not found below certain sizes or above certain densities, given by  \cite{Cappellari2013b} Eq. (4). We also show the mean fit (following the double power law form) for local galaxies in these projections \citep[][Eq. (5)]{Cappellari2013b}. We converted between the two projections of the relations from \cite{Cappellari2013b} using a version of the scalar virial relation $M = 5.0\sigma_e^2R_e/G$ from \cite{Cappellari2006}. For the relations from \cite{Cappellari2013b}, we rescaled $R_e$ to circularized values using the median axis ratio of Coma ($\sim 0.65$), as done in \cite{Beifiori2017}. Here we show a linear fit to the galaxies in JKCS 041 on the $M$-$\sigma_e$ and $M$-$R_e$ planes using the fixed slopes found for the Coma sample on these projections, as derived in \cite{Beifiori2017}. As can be seen from the points in our sample, they extend beyond the zone of exclusion, as is expected for dispersion and size evolution. 

To understand how the galaxies in our sample have evolved from $z \sim 0$ to $z \sim 1.80$, we compared our sample to the Coma cluster (as before with the FP), using the sample compiled and used in \cite{Beifiori2017}, we refer the reader there for more details. To study the evolution of scaling relations, we followed the method of \cite{Newman2012, Cimatti2012, Delaye2014, vanderWel2014, Beifiori2017, Chan2017}. In order to compare the size and velocity dispersion evolution of samples with different mass distributions, the correlation between the two parameters must be removed. To do this, we normalized $\sigma_e$ and $R_{e,B}$ by a mass of $\log(M/M_{\odot}) =11$ for both our mass estimates ($M_{dyn}$ and $M_*^{tot}$) using the relations:
\begin{equation} \label{eq:Rmn}
R_{e,\rm MN} = \frac{R_{e,B}}{(M/10^{11}M_{\odot})^{\beta_{M\text{-}R_e}}},
\end{equation}
\begin{equation} \label{eq:Smn}
\sigma_{e,\rm MN} = \frac{\sigma_{e}}{(M/10^{11}M_{\odot})^{\beta_{M\text{-}\sigma_e}}}.
\end{equation}
Where $R_{e,\rm NM}$ and $\sigma_{e,\rm MN}$ are mass-normalized $R_e$ and $\sigma_e$, $M$ is the mass (either $M_{dyn}$ or $M_*^{tot}$), and the $\beta$ values are the local slopes of Coma as derived from the respective $M$-$R_e$ and $M$-$\sigma_e$ planes \citep{Beifiori2017}.

Using the Coma sample as the local comparison, we then derived the evolution of these mass-normalized structural parameters as a function of redshift. To best address the issue of `progenitor bias' \citep{vanDokkum1996} and attempt to match the JKCS 041 sample to the Coma sample, we selected all galaxies with ages $> 10$ Gyrs from Coma, leaving a sample of five (following the approach of \citealt{Beifiori2017, Chan2016, Chan2017}). Accurately comparing samples of galaxies between different redshifts requires careful comparison of a number properties to best link their evolutionary paths. Although taking an age cut improves this matching of high-redshift galaxies to their possible descendants, this cut was fairly limiting as we do not have ages for all the galaxies in the Coma sample. Refining the comparison further, for example by matching the data to models, was not feasible on the remaining small sample of five galaxies, so we were unable to more accurately address progenitor bias and this may affect the relations derived to quantify structural evolutionary effects on the shift of the FP zero point.

We derived slopes ($\nu$ and $\mu$) from the mass-normalized relations $\Delta \log R_{e,\rm MN} \propto \nu \log(1+z)$ and $\Delta \log \sigma_{e,\rm MN} \propto \mu \log(1+z)$. These values and their respective relations are summarized in Table \ref{tab:evo}, their 1$\sigma$ errors are from bootstrapping. We compared the weighted-mean mass-normalized sizes and dispersions to those of Coma, and found that $M_{dyn}$-normalized sizes of JKCS 041 galaxies were $\sim 69 \%$ smaller, and $M^{tot}_{*}$-normalized sizes were $\sim 74 \%$ smaller than Coma. For the mass-normalized $\sigma_e$ values, we found $M_{dyn}$-normalized dispersions were $\sim 25 \%$ larger in JKCS 041 than in Coma, and for $M^{tot}_{*}$-normalized dispersions we found them to be $\sim 19 \%$ smaller than in Coma.

\subsubsection{Effects of structural evolution on FP ages}
\label{sec:evoage}

We summarize the derived slopes necessary to determine the contribution of structural evolution to the change in FP zero-point in Table \ref{tab:evo}. From Eq. \ref{eq:Lstrevo}, we derived the contribution to the evolution of the FP of just the structural evolution of the size and dispersion using $\chi_{\rm SE}=\left( \frac{2b+1}{b}\nu - \frac{a}{b}\mu\right)$. We determined $\chi_{\rm SE}$ of $-0.06\pm0.42$ and $-0.80\pm0.28$ for $M_{dyn}$-normalized and $M^{tot}_{*}$-normalized parameters respectively. When comparing to an FP zero-point that evolves entirely due to an aging stellar population ($\eta$), we found that the effects of structural evolution may contribute between $\sim 3\%$ and up to $\sim 42\%$ from the $M_{dyn}$ and $M^{tot}_{*}$ normalizations respectively. If we did not apply an age cut to the local Coma sample, this contribution from structural evolution becomes $\sim 2\%$ and $\sim 50\%$ from the $M_{dyn}$- and $M^{tot}_{*}$-normalized parameters respectively. As a comparison of these effects, \cite{Beifiori2017} found comparable contributions of structural evolution to the FP zero-point shift of $\sim 6\%$ and $\sim 35\%$ for $M_{dyn}$ and $M_*$ normalizations respectively.

To test what effect this had on the age values we derived, we used the percentage difference between the slopes derived for luminosity evolution depending entirely on an aging stellar population ($\eta$) and one accounting for structural evolution of galaxies ($\chi$) in Table \ref{tab:evo}. This then translated to a percentage difference in $\Delta \log(M/L)$ which we applied to our sample of seven galaxies using both the $M_{dyn}$-normalized and $M^{tot}_{*}$-normalized slopes. 

For the large structural evolutionary effects we derived for the $M^{tot}_{*}$-normalized parameters ($\sim$ 42$\%$), we found this translated to ages older than the Universe, as also found for XMM2235 in \cite{Beifiori2017}. We therefore capped this maximum age to that of the Universe at $z \sim 1.80$, meaning our derived FP age could be larger by a factor $\sim$ 2.5 when accounting for structural evolution as derived from $M^{tot}_{*}$-normalized parameters. This significant increase could imply that the structural evolution is overestimated for stellar-mass normalized values of $R_{e,B}$ and $\sigma_e$. As \cite{Beifiori2017} suggested, this could be due to a stronger progenitor bias when normalizing by $M^{tot}_{*}$. Given the range of properties of the Coma galaxies, our comparison sample of five old ($> 10$ Gyrs) galaxies may not be ideal descendant matches to our sample from JKCS 041 \citep{vanDokkum2001, Valentinuzzi2010a, Saglia2010, Poggianti2013, Carollo2013, Beifiori2014}. However, with our limited sample size, we were unable to address the effects of progenitor bias further than this. We estimated that the minimal evolutionary effects of the $M_{dyn}$-normalized structural parameters ($\sim 3\%$) on our derived FP ages leads to an increase of only $\sim$ 0.2 Gyrs, which is within the errors.

\begin{figure*} 
\centering
\includegraphics[width=0.7\textwidth , trim=30 10 10 50,clip]{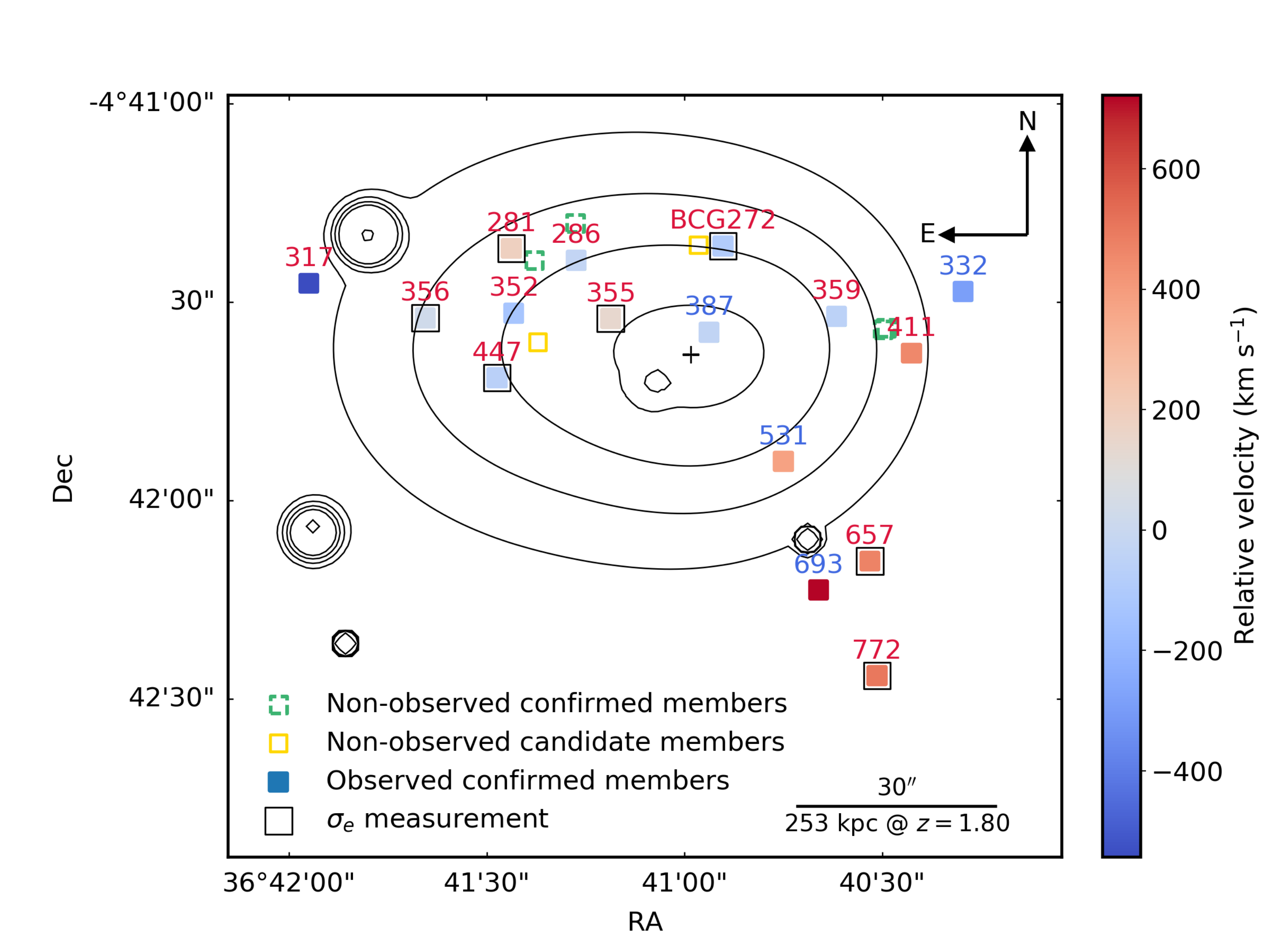}
\caption[]{Spatial extent of the overdensity members with the relative velocity of the 16 observed members indicated. The observed galaxies are marked as quiescent (red IDs) or SF (blue IDs). The non-observed confirmed members (dashed green squares), and non-observed candidate members (yellow squares) are all quiescent. The X-ray contours from \cite{Andreon2009} are shown, and the geometric center of the galaxies is shown by the `+'. The galaxies to the south west of the overdensity show systematically higher relative velocities. See Sec. \ref{sec:struct}.\small}
\label{fig:vel}
\end{figure*}

\begin{figure*}
\includegraphics[width=0.49\textwidth , trim=0 0 50 40 ,clip]{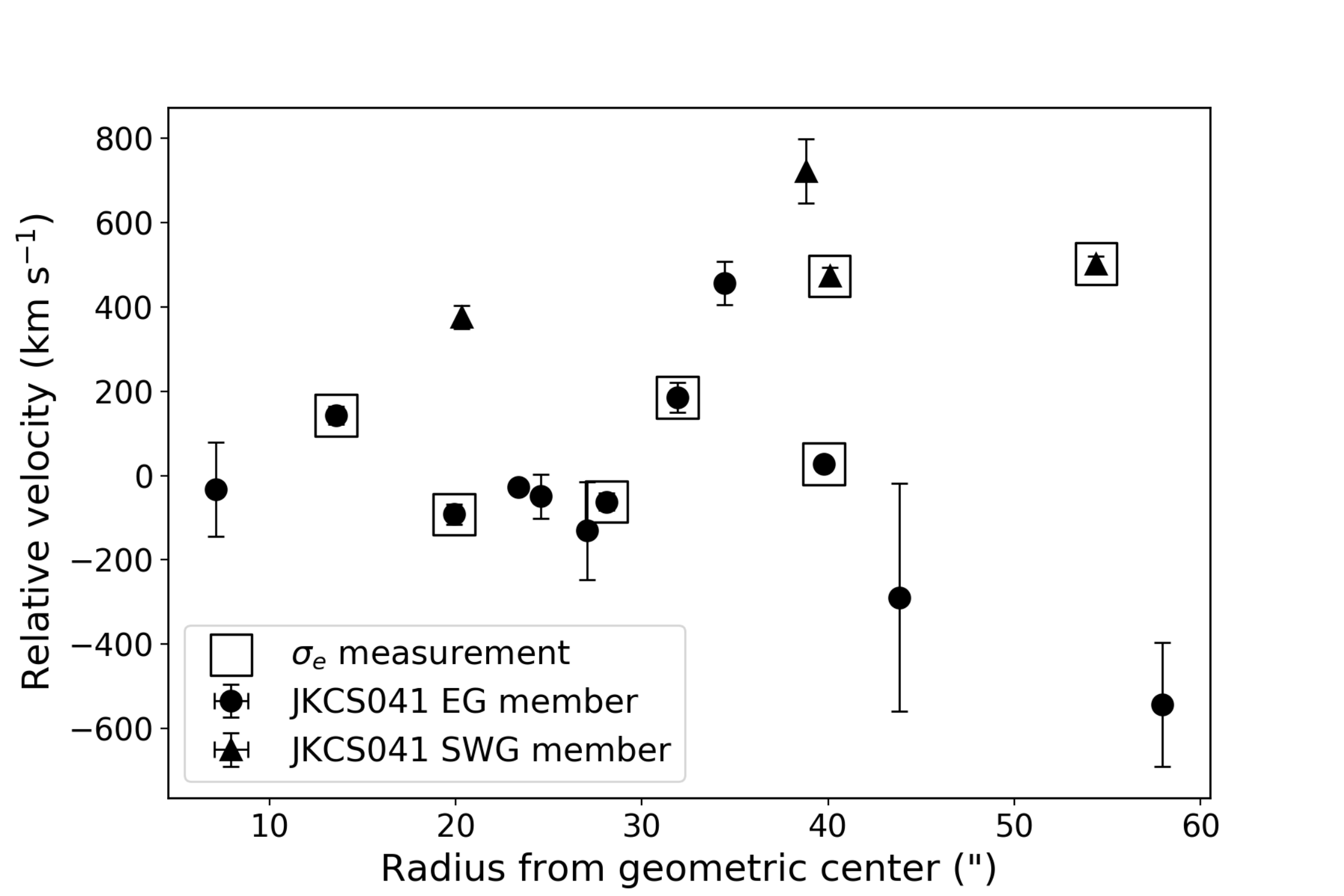}
\includegraphics[width=0.49\textwidth , trim=0 0 50 40,clip]{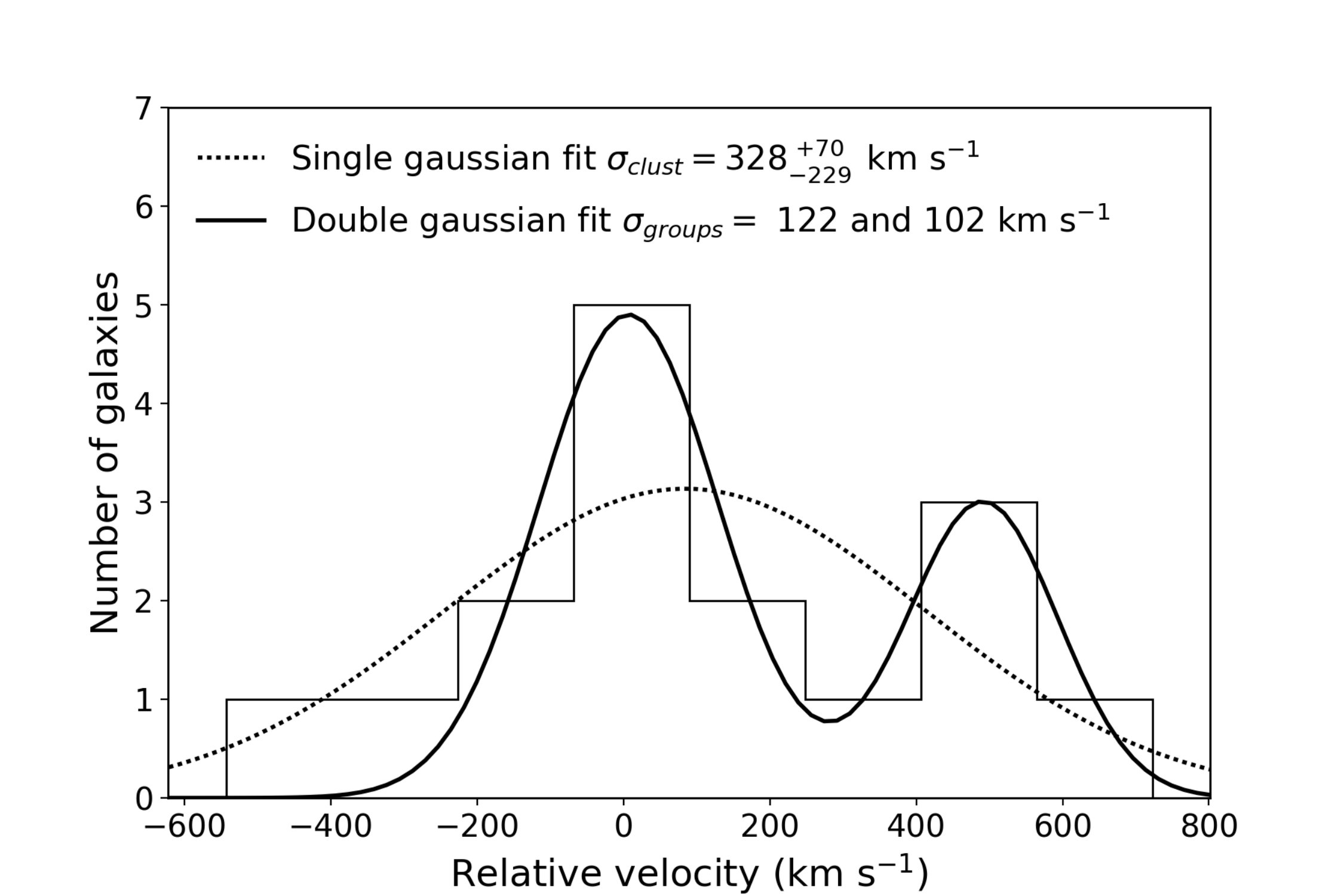}	
\caption[]{Velocities of the galaxies in JKCS 041 relative to the median redshift of the observed sample. \textit{Left:} Phase-space diagram showing all 16 observed galaxies as a function of radius from their geometric center (shown by the `+' in Fig. \ref{fig:vel}). We show the four galaxies in the SWG (triangles; identified in 2D from Fig. \ref{fig:vel}), the EG galaxies (circles), and galaxies with $\sigma_e$ measurements (unfilled squares). \textit{Right:} Histogram of relative velocities of the 16 observed galaxies. We show the best-fitting double Gaussian (solid line, and corresponding dispersions measured from the two peaks), and the single Gaussian fit (dotted line), from which we estimated a $\sigma_{clust}$. See Sec. \ref{sec:dyn}.\small}
\label{fig:dyn}
\end{figure*}

\section{Overdensity structure, dynamics, and ages in 3D}
\label{sec:struct}

\subsection{Structure of JKCS 041}
\label{sec:substruct}

\cite{Newman2014} confirmed 19 (two from emission lines, 17 from continuum and photometric data) members of JKCS 041 and determined \textit{HST} grism redshifts for a further 79 objects (61 from emission lines and 18 from continuum). We have marked all those 79 objects with redshifts that are confirmed non-members of JKCS 041 in Fig. \ref{fig:hst} \citep[small red circles; see][Appendix A]{Newman2014}. Unfortunately, the photo-$z$ catalogue used by \cite{Newman2014} is not public, and the coordinates of 16 non-emission-line objects with 1.4 $< z_{phot} <$ 3 and no grism redshifts are not known. Without these objects, we are unable to draw strong conclusions about the 3D structure of the overdensity. 

As can be seen from the confirmed (green squares) and candidate members (yellow squares) of the overdensity in Fig. \ref{fig:hst}, JKCS 041 appears to be elongated and seems to extend in two distinct directions. In Fig. \ref{fig:vel}, we show the spatial extent of the 16 observed galaxies (filled squares), again with the quiescent (red ID labels) and SF (blue ID labels) members indicated. As a third dimension, we show the velocity relative to the median redshift of the overdensity (blue to red) of the observed galaxies. We also show the confirmed members of JKCS 041 with no KMOS observations (dashed green squares) and unobserved candidate members (unfilled yellow squares), all of which are quiescent. Those galaxies for which we have velocity dispersion measurements are highlighted (unfilled black squares).
 
The observed confirmed members of JKCS 041 that are in the group that extends east (EG) are mostly quiescent, while the group toward the south west (SWG) are 50$\%$ SF galaxies, however this is only four galaxies in total (IDs 531, 657, 693, 772). The X-ray contours overlaid in Fig. \ref{fig:vel} \citep[from][]{Andreon2009} show that the hot diffuse intracluster medium (ICM), is elongated along the EG. In a relaxed cluster, the hot ICM and spatial distribution of galaxies should both approximately trace the potential well of the overdensity. The fact that the X-ray component is offset from the observed galaxies in the SWG could imply that it is not relaxed, that we are not probing the full distribution of members, or that the centering of the contours is incorrect. The image contains a number of bright X-ray point sources, which, if any un-subtracted emission remains, could bias the position of the X-ray contour centroid. 

\subsection{Overdensity dynamics}
\label{sec:dyn}

Galaxies in the SWG show a slight trend of a systematically higher positive relative velocity (Fig. \ref{fig:vel}) than the other observed galaxies. To investigate this offset, we plotted a phase-space diagram and histogram of the relative velocities of the observed galaxies in Fig. \ref{fig:dyn}. In the left panel, we show the relative velocities of the galaxies (from the median measured redshift), as a function of their radius from the geometric center of the galaxies (average in RA and Dec, marked by the `+' symbol in Fig. \ref{fig:vel}). We show the four galaxies in the SWG (IDs 531, 657, 693, 772; triangles), as identified in 2D on Fig. \ref{fig:vel}, the EG galaxies (circles), and galaxies for which we derived $\sigma_e$ measurements (unfilled squares). We see that the four galaxies  occupy a distinct region of the phase-space diagram at higher relative velocities and larger radii for three of the galaxies, with one interloper (ID 447), which in 2D space is close to the SWG.

To investigate the dynamics of the cluster further, we show a histogram of the relative velocities of the galaxies in the right panel of Fig. \ref{fig:dyn}. We fitted the distribution of velocities with both a double and single Gaussian and found that it was best fitted by a double Gaussian. Although this implies the overdensity is not virialized, we used the histogram of galaxy velocities to investigate how a derived virial mass compared to other mass estimates for this overdensity. The virial mass of a relaxed overdensity, originally presented by \cite{Limber1960} and modified by \cite{Carlberg1996}, is given by
\begin{equation} \label{eq:Mvclust}
M_v = \frac{3\pi}{2} \frac{\sigma_{clust}^2R_h}{G}.
\end{equation}
Where $R_h$ is the ring-wise projected harmonic mean radius, defined by
\begin{equation} \label{eq:Rh}
\begin{aligned}
R_h = & \frac{N(N-1)}{\sum_{i<j}\frac{1}{2\pi}\int^{2\pi}_{0}\frac{d\theta}{\sqrt{R_i^2 + R_j^2 + 2R_iR_i\cos{\theta}}}}\\
= & \frac{N(N-1)}{\sum_{i<j}\frac{2\pi}{R_i + R_j}K(k_{ij})}.
\end{aligned}
\end{equation}
Here $R_i$ and $R_j$ are the distances of galaxies $i$ an $j$ from the central point of the overdensity (which we have defined as the geometric center of the galaxies; `+' in Fig. \ref{fig:vel}), $k_{ij}^2=4R_iR_j/(R_i+ R_j)^2$, and  $K(k_{ij})$ is the complete elliptical integral of the first kind in Legendre's notation \citep[form from][]{Irgens2002}. 

For all 20 confirmed members of JKCS 041, we derived $R_h = 493\pm20$ kpc. For the 16 observed galaxies, we obtain $\sigma_{clust} = 328\substack{+71\\-229}$ km s$^{-1}$ from a single Gaussian fit to the galaxies (Fig. \ref{fig:dyn}), and 1$\sigma$ errors from randomly sampling the observed galaxies and refitting. This gives a poorly constrained total mass for the overdensity $\log(M_{clust}/M_{\odot}) = 13.8\pm0.6$. Within errors, this is consistent with the value found by \cite{Andreon2014} of $\log(M/M_{\odot})\gtrsim14.2$. However, as discussed, it appears the observed galaxies from which we derived a $\sigma_{clust}$ value form two dynamically distinct groups, meaning that the overdensity is not virialized, making this an unreliable measure of its total mass.

\subsection{Galaxy ages}
\label{sec:galage}

To further test the properties of the galaxies in the two spatially and dynamically distinct groups, we investigated the ages we could derive from the seven galaxies on the FP that reside in the EG and SWG separately (indicated in Figs. \ref{fig:fp} and \ref{fig:MtL}). With our seven galaxies in total, we realize we are dealing with very small numbers and splitting the sample in this way based on their location in the overdensity causes large uncertainties. Deriving ages from the FP as described above, we took the two SWG galaxies that were on the FP (IDs 657 $\&$ 772) and the five that were in the EG, and determined ages for the two groups. We found that the EG galaxies had a significantly older mean age $2.1\substack{+0.3 \\ -0.2}$ Gyrs ($z_{form} = 4.1\substack{+0.7 \\ -0.4}$), than the two SWG galaxies with mean age $0.3\pm0.2$ Gyrs ($z_{form} = 2.0\substack{+0.2 \\ -0.1}$). To check this was not a direct effect of mass, we determined average light-weighted stellar masses of the EG ($\langle \log(M^{tot}_*/M_{\odot}) \rangle = 11.36$) and SWG ($\langle \log(M^{tot}_*/M_{\odot}) \rangle = 11.18$) populations, and found that these populations had similar masses. However, if we look just at the dynamical masses then this age difference may also be attributed to the fact that the SWG galaxies have lower dynamical masses.

%------------------
%DISCUSSION
%------------------
\section{Discussion}
\label{sec:disc}

\subsection{Absorption line spectroscopy at $z \gtrsim 1.80$}
\label{sec:discspec}

For JKCS 041 at $z \simeq 1.80$, we have derived seven $\sigma_e$ values for individual galaxies. In the literature, dispersion measurements of 16 individual galaxies at $z \gtrsim 1.8$ have been confirmed \citep{vanDokkum2009, Onodera2010, Toft2012, vdSande2013, Belli2014b, Belli2017, Newman2015, Hill2016, Toft2017}. Most of these observations were of the brightest, most massive galaxies spanning a wide range of redshifts. These seven stellar velocity dispersions for galaxies in JKCS 041 increases the sample in the literature by $>$ 40$\%$. We also present the largest number of galaxy velocity dispersions within a single overdensity at comparable redshifts, as no previous survey has systematically targeted quiescent galaxies to these depths and distances.

\subsection{Comparison of FP ages for JKCS 041}
\label{sec:discfp}

The mean age obtained for the seven JKCS041 galaxies on the FP is $1.4\pm0.2$ Gyrs ($z_{form} = 3.0\pm0.3$), this is consistent within 1$\sigma$ errors of previous results for this overdensity \citep{Newman2014, Andreon2014}. \cite{Newman2014} obtained ages from spectral fitting of stellar absorption lines of stacked \textit{HST} grism spectra of the 15 quiescent galaxies. They fitted the galaxies in two mass bins, for the high mass ($\log(M_*/M_{\odot}) >$ 11) quiescent members, their models gave an age of $1.45\substack{+0.24 \\ -0.18}$ Gyrs ($z_{form} = 3.0\substack{+0.4 \\ -0.2}$) and for the lower mass sample $10.5 < \log(M_*/M_{\odot}) < 11$, they obtained a luminosity-weighted age of $0.90\substack{+0.19 \\ -0.10}$ Gyrs ($z_{form} = 2.4\substack{+0.2 \\ -0.1}$). Using the same \textit{HST} grism data, \cite{Andreon2014} derived SFH weighted ages and found average ages for average masses in these bins of 1.4 Gyrs at $\log(M_*/M_{\odot}) \sim$ 11.5, and 0.7 Gyrs at $\log(M_*/M_{\odot}) \sim$ 10.5, and found a mean age of all the galaxies of 1.1 $\pm$ 0.1 Gyrs for mean mass at $\log(M_*/M_{\odot}) \sim$ 11.

\subsection{Comparison of FP ages to other KCS overdensities}
\label{sec:compkcs}

The mean formation redshift we derived for the seven galaxies on the FP in JKCS 041 ($z_{form} = 3.0\pm0.3$), is consistent within errors of just the five high-mass $\log(M_*/M_{\odot}) >$ 11 galaxies ($z_{form} = 2.8\substack{+0.5 \\ -0.4}$). This formation epoch is consistent with that determined for the richest and most virialized KCS overdensity analyzed in \cite{Beifiori2017}; XMM2235 at $z \sim 1.39$ ($z_{form} = 2.95$). \cite{Beifiori2017} found that XMM2235 had a slightly older formation epoch than the other two clusters, which they suggested might indicate more rapid evolution for this more massive and relaxed cluster. JKCS 041, one of the most massive overdensities at $z \sim$ 1.8 and with a high passive fraction, has the same formation epoch. This adds weight to the indicative results found by \cite{Beifiori2017}, that they concluded might imply some accelerated formation in more massive overdensities at earlier times \citep[as also found by e.g.,][]{Gebhardt2003, Saglia2010}. As can also be seen from the total cluster mass versus redshift plot for the KCS sample in Fig. 1 of \cite{Beifiori2017}, XMM2235 and JKCS 041 lie on the same model mass-accretion evolutionary track of a massive local cluster ($\log(M_{200}/$h$^{-1}$M$_{\odot}) = 16$). This also implies that JKCS 041 and XMM2235 could have similar cluster evolution.

\subsection{Structure and evolution of JKCS 041}
\label{sec:discstruct}

When looking at just the 16 confirmed members of JKCS 041 that we observed with KMOS, we found that the EG contained the oldest galaxies (as measured from the FP of five galaxies), while the younger SWG (as determined from the FP ages of two galaxies) contained a higher proportion of SF galaxies. The age results for the two groups of galaxies, although based on very small numbers of galaxies, were found to be independent of light-weighted stellar mass. The two younger SWG quiescent galaxies, have very young ages that were giving us a significantly younger mean age for all seven galaxies on the FP. Although the mean age we derived for all seven galaxies is consistent with a more massive and virialized overdensity, these results hint that JKCS 041 could be made up of two different age populations. However, for an age derived for only two points, this result has low significance. 

As can be seen from Fig. 8, the SWG galaxies have smaller $M_{dyn}$ values than the other five galaxies in the dispersion sample, but comparable light-weighted stellar masses. The FP age results indicate that the SWG galaxies are significantly younger than the EG galaxies. Trends seen in the field indicate that lower mass ETGs have younger ages \citep{Treu2005a, Thomas2010}, which is more consistent with the $M_{dyn}$ values. However, for the light-weighted stellar masses, although there is significant scatter in the NMBS relation that they were derived from, their errors reflect this uncertainty and we find them to be consistent with those \cite{Newman2014} derived from SED fits. We note that the mass estimates are not in agreement (within errors) for only three of our seven galaxies, and so the apparent difference in mass estimates has low significance. The possible difference between the two mass determinations may be explained by the different observational limitations associated with each. However, with the current limited sample of dynamical masses for individual galaxies at this redshift, these effects are hard to quantify.

From the overdensity dynamics, we see two distinct groups of galaxies, separated both spatially and in relative velocity, with the SWG having systematically higher relative velocities. Given the dynamics of the 16 observed galaxies and age results (of the seven galaxies on the FP) of the confirmed members of JKCS 041, we suggest that these results could hint at an overdensity in formation that could be made up of two distinct merging groups of galaxies. This tentative result might be consistent with the work done by \cite{Newman2014} who showed a mass matched field sample \cite[from][]{Whitaker2013} at the same redshift as JKCS 041 were not as quiescent, and that environmental quenching was responsible for the increased number of quiescent galaxies in JKCS 041. This result could also be interpreted as the increased number of quiescent galaxies in the EG which we found to be older and potentially form a more relaxed group of the overdensity, and the lower proportion of quiescent galaxies in the SWG could potentially be a merging group that represents a transition population between typical field and overdensity populations at this redshift. This is also consistent with the difference in ages of the two groups, as it has been found that ETGs in the field contain younger stellar populations than cluster ETGs \citep{vanDokkum2003}. However, strong conclusions cannot be drawn about the evolutionary scenario of the galaxies based on the limited sample.

%--------------------
%CONCLUSION
%--------------------
\section{Conclusions}
\label{sec:conc}

In this paper we present new KMOS spectra for the overdensity JKCS 041 at $z = 1.80$ as part of KCS. KCS is a GTO KMOS program, which aimed to constrain the evolution of galaxies in dense environments between 1 $< z <$ 2. \cite{Newman2014} confirmed 19 members of JKCS 041 using \textit{HST} grism spectra, and identified three candidate members. We observed 16 galaxies with KMOS (12 quiescent and four SF), 15 confirmed members and one candidate member (ID 772), which we subsequently spectroscopically confirmed, bringing the total number of confirmed members of JKCS 041 to 20. 

We reduced and analyzed \textit{HST} images of the overdensity in the $H_{160}$ and $Y_{105}$ bands (presented in \citealt{Newman2014}). To determine photometric parameters of the galaxies, we fitted 2D S{\'e}rsic profiles to galaxies using \textsc{galapagos-2.2.5b}. From the spectra, we determined improved spectroscopic redshift measurements for the 16 observed galaxies using either kinematic fits (for 14 galaxies using \textsc{ppxf}), or emission lines (in the case of two SF galaxies). From kinematic fits of the quiescent galaxies in our sample, we were able to determine stellar velocity dispersions for seven galaxies. We combined these dispersions with the photometric parameters to construct a fundamental plane of individual galaxies in JKCS 041. We then used the FP to derive ages of galaxies. With the improved redshifts, we were able to investigate the three-dimensional dynamics of observed galaxies in the overdensity. The main results of this work are summarized below.

\begin{enumerate}
\item Using photometric parameters derived from the \textit{HST} images, and the derived $\sigma_e$ values, we were able to construct an FP for seven quiescent galaxies in JKCS 041. This is the highest redshift FP constructed for a single overdensity. It further supports studies suggesting that the FP holds to $z \sim 2$ \citep[e.g.][]{Bezanson2013b}. 

\item From the shift in zero-point of the FP, we estimated $M/L$ evolution with $z$ for the galaxies in JKCS 041. Overlaying derived $\Delta \log(M/L_B)$ evolutionary tracks from interpolated SSP \citep{Maraston2005} models, we derived a mean age of the seven galaxies to be $1.4\pm0.2$ Gyrs ($z_{form} = 3.0\pm0.3$). Comparisons with the literature showed that these results were consistent with other studies of JKCS 041, and results of the other KCS overdensities \citep{Beifiori2017}. 

\item Testing the effects of structural and stellar velocity dispersion evolution on these values, we found very little effect when using $M_{dyn}$-normalized parameters ($\sim$ 0.2 Gyrs), but up to a factor of $\sim 2.5$ larger ages when using $M^{tot}_{*}$-normalized parameters. The large difference between the effects from different mass normalizations could mean an overestimation of structural evolution from $M^{tot}_{*}$-normalized values, this could be due to progenitor bias.

\item From the dynamics of 16 confirmed members of JKCS 041, we see a distinct group of galaxies extending south west in the overdensity. These few galaxies, with a higher SF proportion, have systematically higher relative velocities. As a further investigation into the structure of JKCS 041, we determined ages for those galaxies on the FP in the east, and south-west groups. We found significantly older ages of the galaxies making up the EG ($2.1\substack{+0.3 \\ -0.2}$ Gyrs, $z_{form} = 4.1\substack{+0.7 \\ -0.4}$), than the two quiescent galaxies in the SWG ($0.3\pm0.2$ Gyrs, $z_{form} = 2.0\substack{+0.2 \\ -0.1}$). These tentative dynamic and age results might indicate the overdensity is in formation and made up of two merging groups of galaxies.
\end{enumerate}

%------------------
%ACKNOWLEDGEMENTS
%------------------
\acknowledgments
\noindent We thank the entire KMOS instrument and commissioning teams for their hard work, which has allowed our observing program to be carried out successfully. We wish to thank the ESO staff, and in particular the staff at Paranal Observatory, for their support during observing runs over which the KMOS GTO observations were carried out. We thank the anonymous referee for their insightful feedback that helped improve the clarity of the paper. LJP wishes to thank Boris H{\"a}u{\ss}ler for useful advice regarding \textsc{galapagos-2.2.5b} and Amelie Saintonge for useful comments. LJP is supported by a Hintze Scholarship, awarded by the Oxford Centre for Astrophysical Surveys, which is funded through generous support from the Hintze Family Charitable Foundation. RLD acknowledges travel and computer grants from Christ Church, Oxford and support from the Oxford Centre for Astrophysical Surveys, which is funded by the Hintze Family Charitable Foundation. JCCC acknowledges the support of the Deutsche Zentrum f{\"u}r Luft- und Raumfahrt (DLR) via Project ID 50OR1513. MC acknowledges support from a Royal Society University Research Fellowship. DJW acknowledges the support of the Deutsche Forschungsgemeinschaft (DFG) via Projects WI 3871/1-1 and WI 3871/1-2.

%------------------
%BIBLIOGRAPHY
%------------------
\bibliography{JKCS041bib}

%------------------
%APPENDICES
%------------------
\appendix
\section{Comparison of derived photometric parameters}
\label{sec:compphot}

\begin{figure*} 
\centering
\includegraphics[width=0.4\textwidth , trim=20 20 50 50,clip]{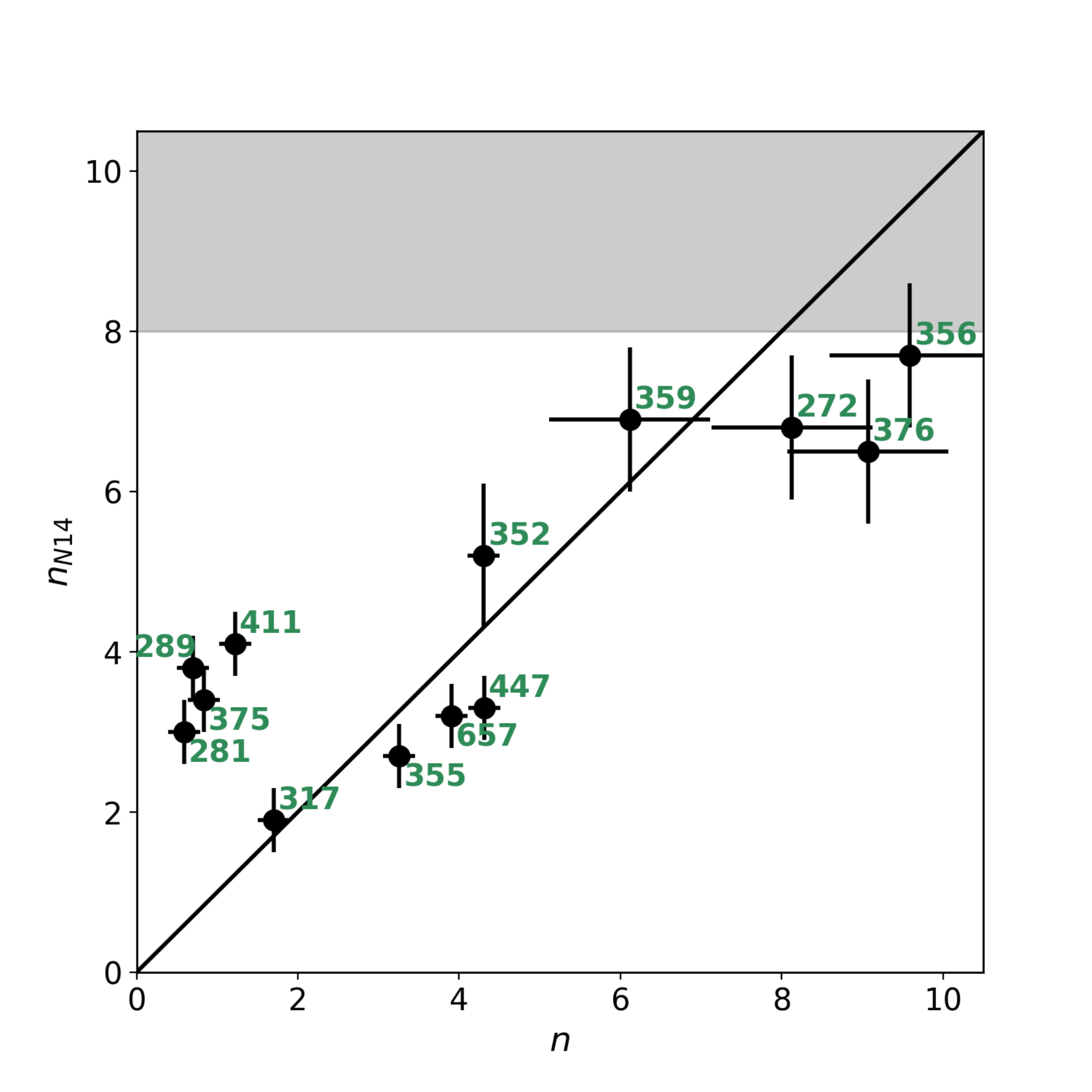}
\caption[]{Comparison of S{\'e}rsic indices derived in this paper ($n$) and those of \cite{Newman2014} ($n_{N14}$) and a one-to-one line for reference. There is a slight trend between the $n$ values we derived and $n_{N14}$. The trend at the high end may be due to the limit imposed by \cite{Newman2014} ($<$ 8; shaded region). The galaxies for which we derived the lowest $n$ values, have systematically higher $n_{N14}$ values. The photometry for these galaxies show that most are very compact (IDs 281, 289, 411), as expected for a low $n$ ($\lesssim 1$) values that we found. The other is in a close pair (ID 375) that we discuss further in App. \ref{sec:compphot}.\small}
\label{fig:n}
\end{figure*}

In this paper we reduced and derived photometric parameters from \textit{HST} images that were presented in \cite{Newman2014}. To test the reliability of our derived photometry, we compared the values derived using \textsc{galapagos-2.2.5b} in this paper to the values published in \cite{Newman2014}. We first compared the S{\'e}rsic indices we derived for the 13 quiescent galaxies for which values were published in \cite{Newman2014} ($n_{N14}$) within their imposed limits of $0.2 < n_{N14} < 8.0$ in Fig. \ref{fig:n}. We see a slight trend between the $n$ values we derived and those of \cite{Newman2014} where our lowest $n$ values tended to be lower than $n_{N14}$ and our highest $n$ values were higher than $n_{N14}$. However, the three largest $n$ values were $>$ 8, which was the limit imposed by \cite{Newman2014} (shaded region), which could explain the trend seen at the highest values. For the lowest $n$ values we derived, the photometry showed that most of these galaxies were very compact (IDs 281, 289, 411). A lower S{\'e}rsic index might be expected for compact galaxies, however the difference can probably be attributed to differences in sky subtraction or estimation. The other is in a close pair (ID 375) that we discuss below.

\begin{figure*} 
	\includegraphics[width=0.5\textwidth, trim=25 10 60 60,clip]{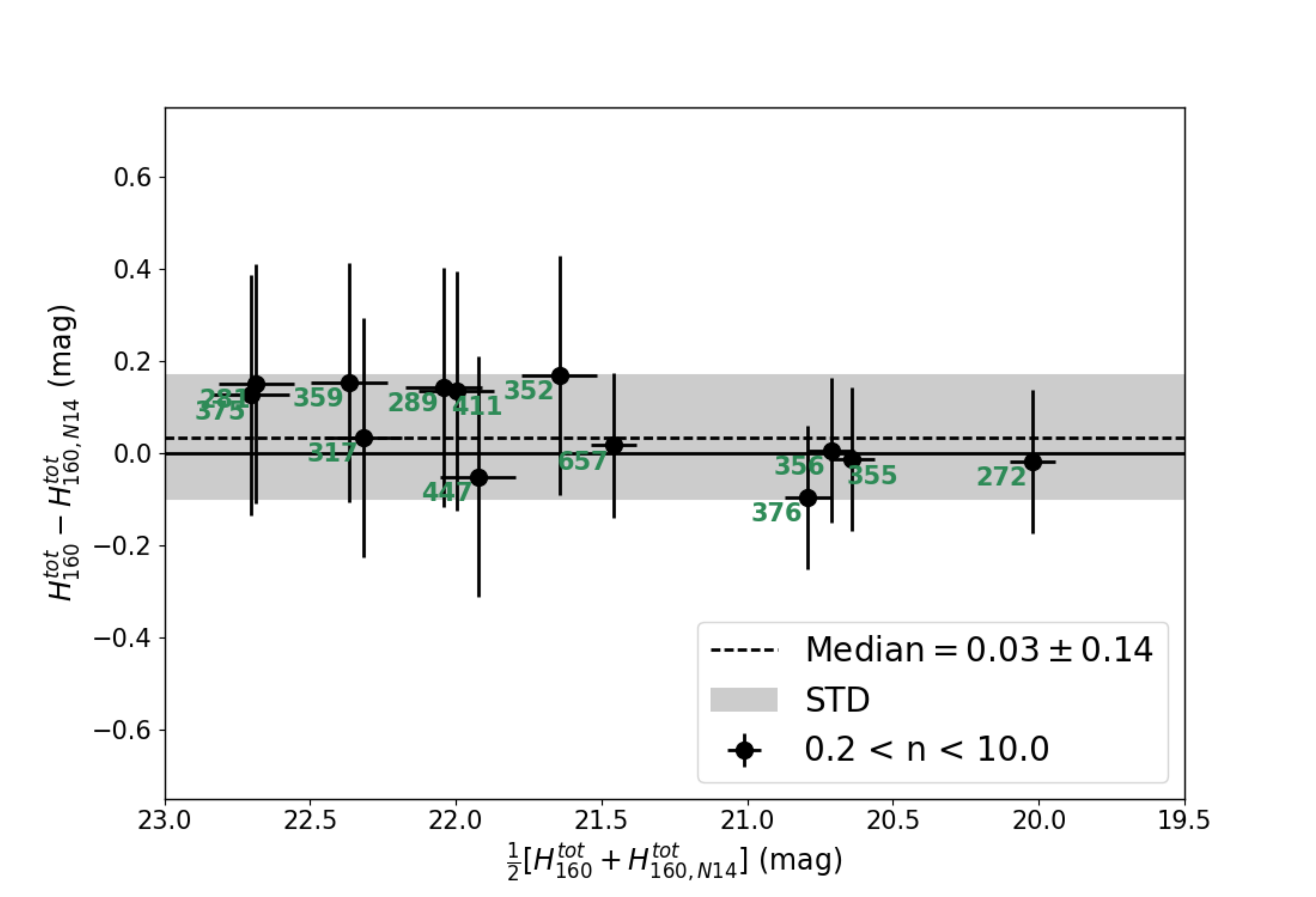}
	\includegraphics[width=0.50\textwidth , trim=25 10 60 40,clip]{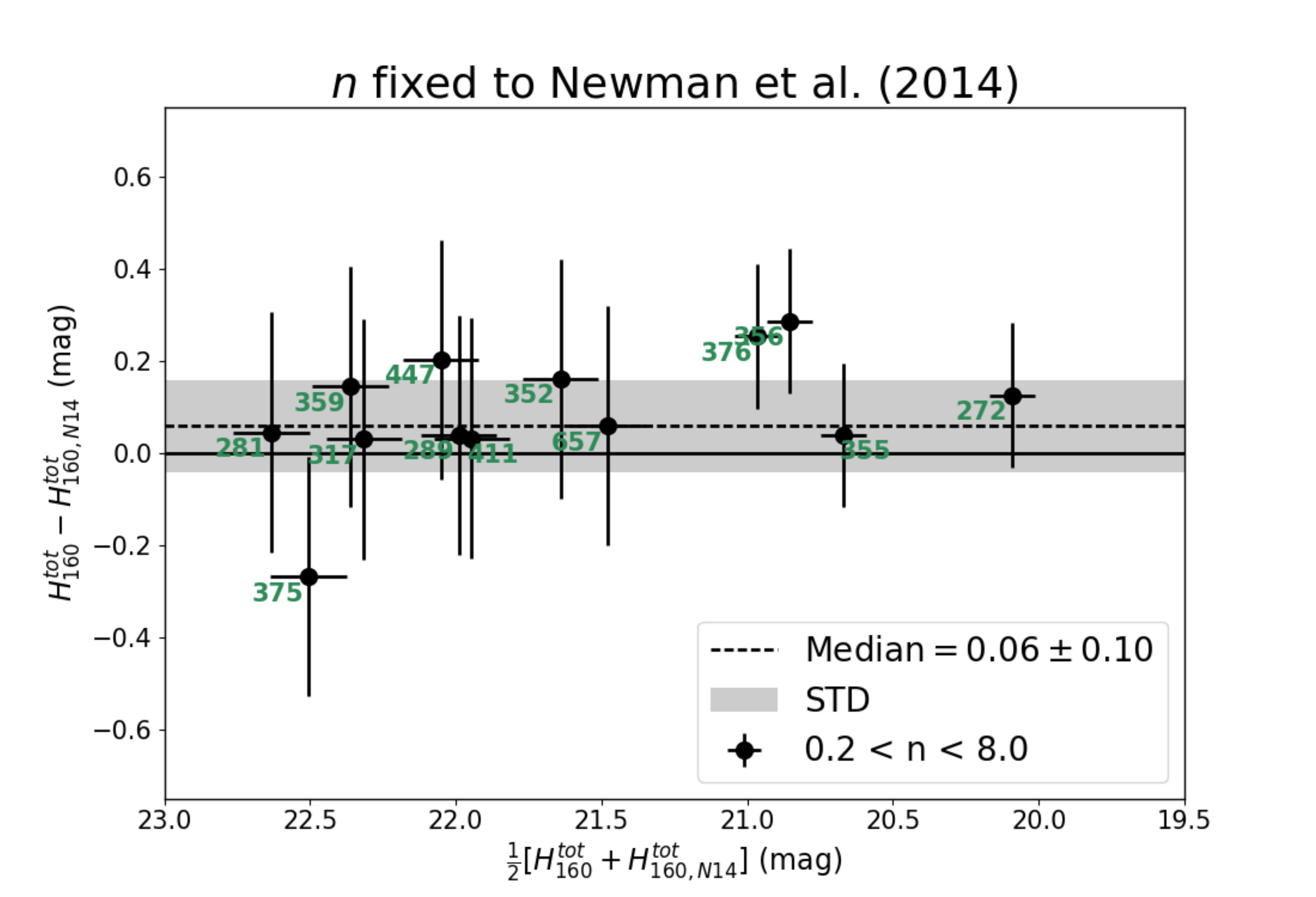}\label{H160comp_fixn}\\
	\includegraphics[width=0.5\textwidth , trim=25 10 60 60 ,clip]{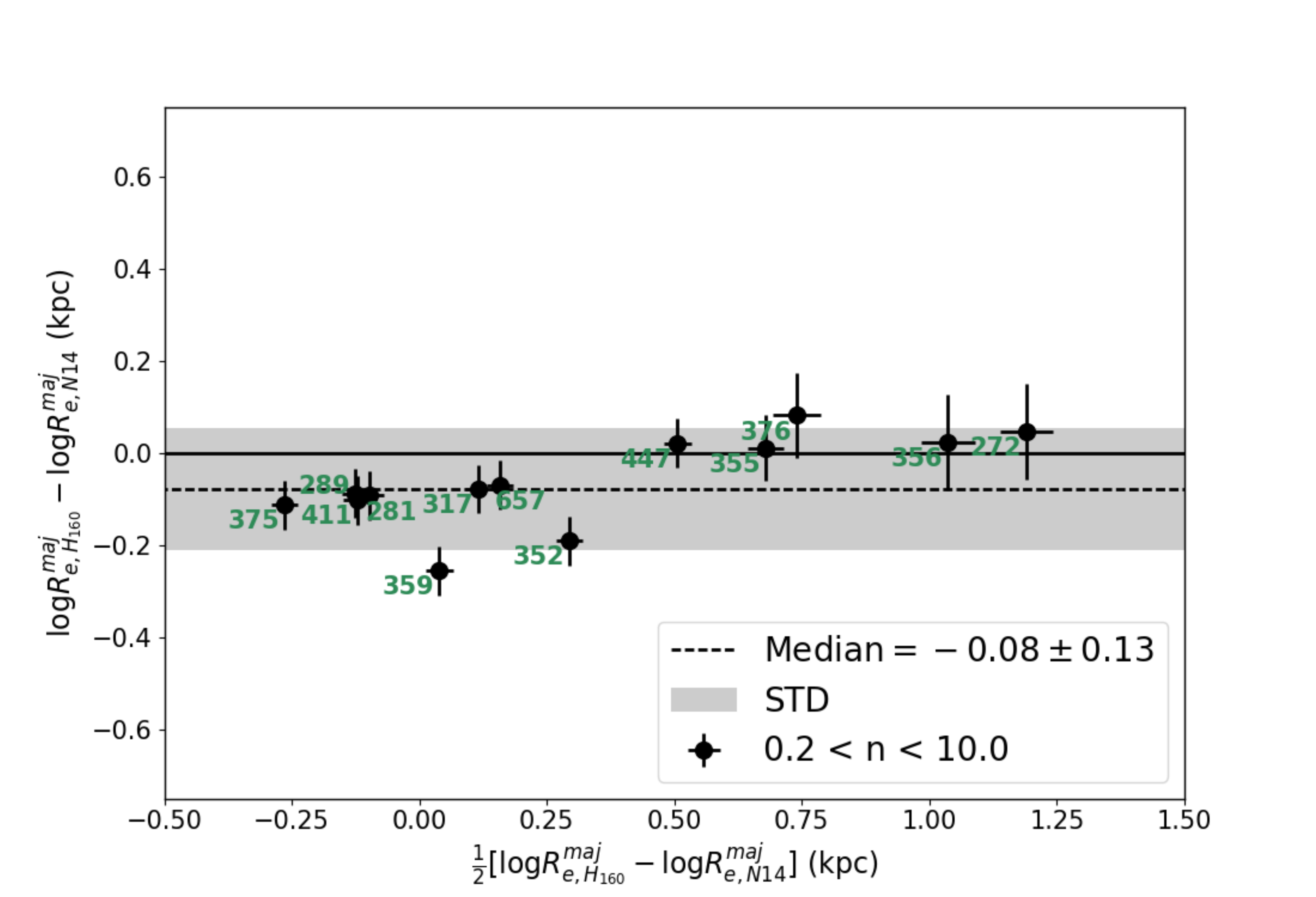}\label{Recomp}
	\includegraphics[width=0.5\textwidth , trim=25 10 60 60,clip]{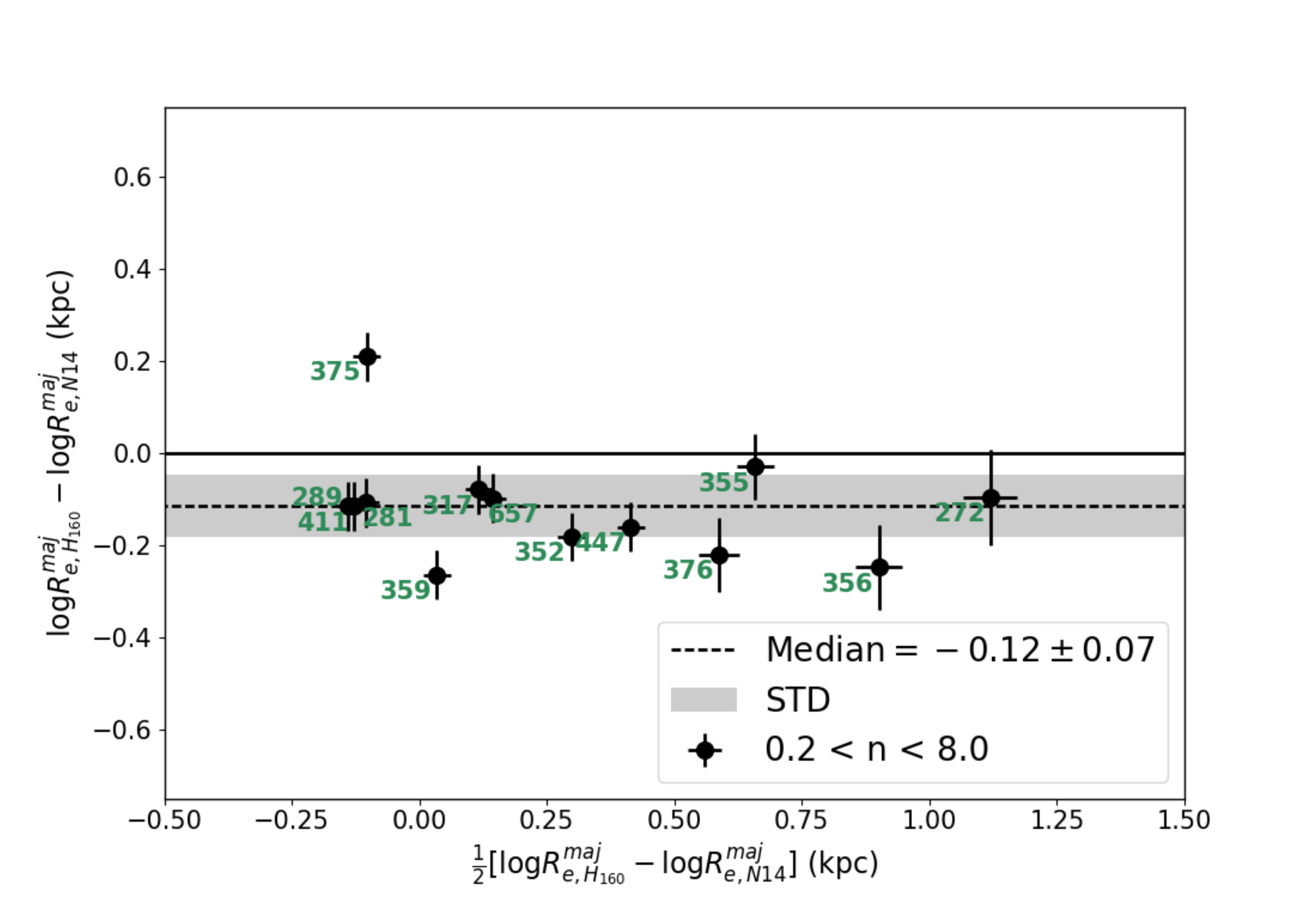}\label{Recomp_fixn}\\
	\includegraphics[width=0.5\textwidth , trim=25 10 60 60 ,clip]{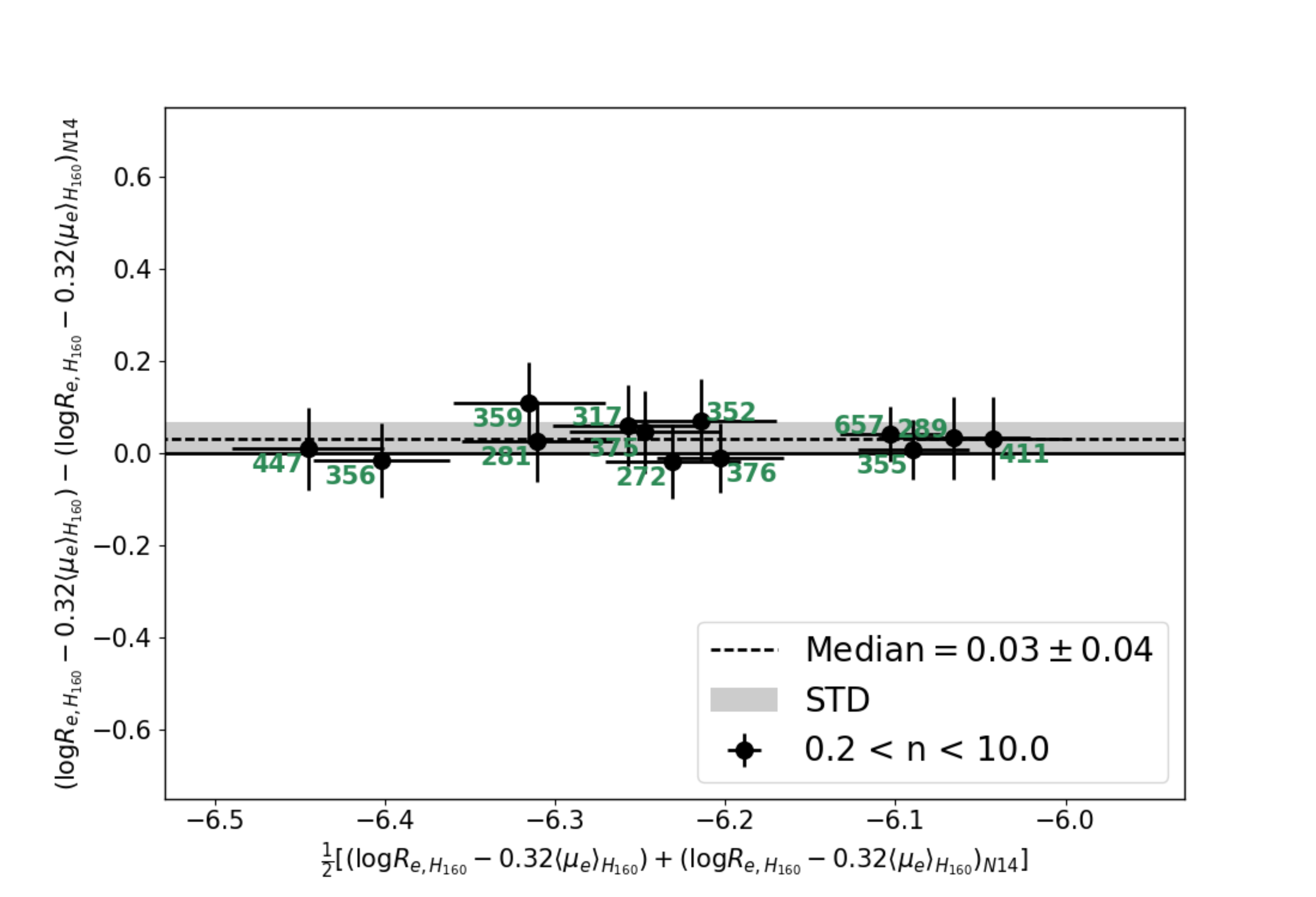}\label{Iecomp}
	\includegraphics[width=0.5\textwidth , trim=25 10 60 60,clip]{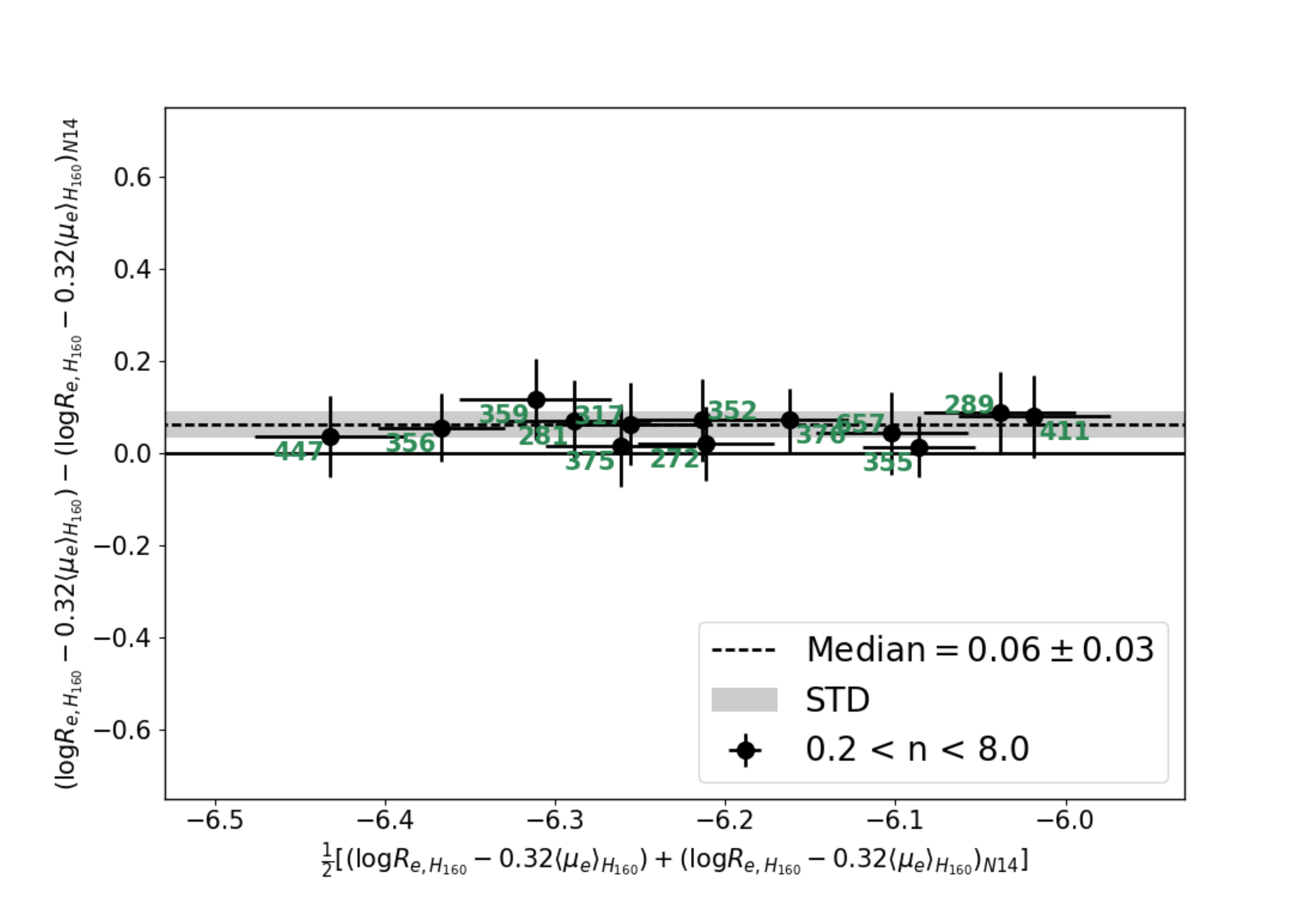}\label{Iecomp_fixn}\\
\caption[]{Comparison of derived photometric parameters between this work and that of \cite{Newman2014} (N14). Both studies derived values from the same \textit{HST} images, but different reduction and analyses. Here we show parameters derived from the $H_{160}$ band. We plot the difference against the average of three parameters for 13 confirmed quiescent galaxies in JKCS 041 that were also fitted by \cite{Newman2014}. We show the galaxy IDs, and compare our derived values on the left, and values derived at fixed-$n$ \citep[of][]{Newman2014} on the right. \textit{Top:} Total integrated magnitude from S{\'e}rsic fits ($H^{tot}_{160}$). \textit{Middle:} $R_e$ along the major axis ($R^{maj}_e$). \textit{Bottom:} Combined photometric parameters from the FP (in the $H_{160}$ band for comparison with values from \citealt{Newman2014}); the circularized $R_{e,H_{160}}$ (kpc) and average surface brightness within a circularized $R_e$ -- $\langle \mu_e \rangle_{H_{160}}$ (mag arcsec$^{-2}$). These are combined using typical coefficients for this definition of the FP ($R_{e,H_{160}} - 0.32 \langle \mu_e \rangle_{H_{160}}$; e.g. \citealt{Bender1998}).\small}
\label{fig:compphot}
\end{figure*}

In Fig. \ref{fig:compphot} is the comparison of our derived photometric parameters  to those of \cite{Newman2014}: total integrated magnitude from S{\'e}rsic fits ($H^{tot}_{160}$), and effective radius along the major axis in the $H_{160}$ band ($R^{maj}_{e,H_{160}}$) in kpc assuming $z=1.80$ in our cosmology. We also show the FP parameter but in the $H_{160}$ band for comparison with \cite{Newman2014}. We combined circularized $R_{e,H_{160}}$ (kpc) and average $H_{160}$ surface brightness within $R_{e,H_{160}}$ ($\langle \mu_e \rangle_{H_{160}}$; mag arcsec$^{-2}$) using typical coefficients for this definition of the FP ($R_{e,H_{160}} - 0.32 \langle \mu_e \rangle_{H_{160}}$; e.g. \citealt{Bender1998}). Shown on these plots is the difference against the average value of the parameters for each galaxy between the two studies (left column). As an additional test, we also show the derived parameters we obtained if we fixed the S{\'e}rsic indices for the galaxies to those in \cite{Newman2014}, and fitted the galaxies using \textsc{galapagos-2.2.5b}. We show the median difference of each parameter between the studies (dashed line) and the 1$\sigma$ errors ($=1.4826\times$MAD) of the distribution (shaded region). We also show the individual galaxy IDs.

In general, we found that the parameters were all consistent within 1$\sigma$ errors. We found slightly fainter $H^{tot}_{160}$ magnitudes (positive difference), smaller $R_e$ values (negative difference), and consistent surface brightnesses. For all parameters, galaxy ID 375 is consistently marginally deviating from the rest of the galaxies. This galaxy is in a close pair with its more massive companion ID 376, so we did not prioritize these galaxies for KMOS observations due to their proximity. Using \textsc{galapagos-2.2.5b} we robustly determined a sky value for each object using a flux-curve growth method, measuring the sky in a series of elliptical annuli and rigorously masking sources. \cite{Newman2014} used a single region bounded by concentric rectangles and masking of sources to determine a sky level, and a smaller fitting region ($2.5 \times$Kron radius of the galaxy, as compared to the $5\times$Kron radius we used). The differences between the sky estimation and fitting region most likely explains the discrepancy between our derived parameters for this galaxy.

%------------------
%END DOC
%------------------
\end{document}